\documentclass{aa}  
\usepackage[varg]{txfonts}
\usepackage{graphicx}
\usepackage{siunitx}
\usepackage{xcolor}
\usepackage{natbib}
\usepackage{soul}
\usepackage{booktabs}
\usepackage{boldline}
\usepackage{nccmath}
\usepackage{amsmath}
\usepackage{tabularx}
\usepackage{algorithm2e}
\usepackage{csquotes}
\usepackage{comment}
\usepackage{longtable}
\usepackage{tabu}
\usepackage{gensymb}
\usepackage{enumitem}
\usepackage{float}
\usepackage{subcaption}
\usepackage[labelfont=bf]{caption}
\usepackage{mwe}
\usepackage[colorlinks,citecolor=blue,urlcolor=blue,bookmarks=false,allcolors=blue,hypertexnames=true]{hyperref} 
\usepackage{threeparttable}
\usepackage{color}
\usepackage[colorinlistoftodos]{todonotes}
\usepackage{tabularx}
\usepackage{adjustbox}
\usepackage{blindtext, rotating}
\usepackage{needspace} 

\usepackage{array,booktabs}

\newcommand\clearrow{\global\let\rowmac\relax}
\makeatletter

\DeclareRobustCommand{\OI}{%
  \mbox[{O\check@mathfonts\fontsize\sf@size\z@\selectfont I}]~%
}
\DeclareRobustCommand{\OII}{%
  \mbox[{O\check@mathfonts\fontsize\sf@size\z@\selectfont II}]~%
}
\DeclareRobustCommand{\OIII}{%
  \mbox[{O\check@mathfonts\fontsize\sf@size\z@\selectfont III}]~%
}
\makeatother
\def \reff       {$R_\textrm{{eff}}$}
\def \width      {$W_{80}$}
\def \oiiiwidth  {$W^\textrm{\OIII}_{80}$}
\def \stwidth   {$W^{\star}_{80}$}
\def \q          {$Q_{W80}$}
\def \kms         {\,km$\,$s$^\mathrm{-1}$}

\def \arcsec      {\text{$^\mathrm{\prime\prime}$}}

\def \whz         {\,W\,Hz$^{-1}$}

\newcommand{\hbeta}{\text{H\textsc{$\beta$}}}
\newcommand{\halpha}{\text{H\textsc{$\alpha$}}}

\newcommand{\oiiihbeta}{\text{[O\textsc{iii}]}/\text{H\textsc{$\beta$}}}

\newcommand{\Nii}{\text{[N\textsc{ii}]}}

\newlist{inlineroman}{enumerate*}{1}
\setlist[inlineroman]{itemjoin*={{ }},afterlabel=~,label=\roman*)}

\makeatletter
\renewcommand{\fnum@figure}{Figure \thefigure}
\makeatother

\begin{document} 
\title{Feedback from low-to-moderate luminosity radio-AGN with MaNGA}
\author{Pranav Kukreti\inst{1}\thanks{\email{kukreti@uni-heidelberg.de}}, Dominika Wylezalek\inst{1}, Marco Alb\'an\inst{1} and Bruno DallAgnol de Oliveira\inst{1}}

\authorrunning{Kukreti et al.}

\institute{Astronomisches Rechen-Institut, Zentrum für Astronomie der Universität Heidelberg, Mönchhofstr. 12-14, 69120 Heidelberg,\\
Germany}

  \abstract{Spatially resolved spectral studies of radio-AGN host galaxies have shown that these systems can impact the ionised gas on galactic scales. However, whether jet and radiation-driven feedback occurs simultaneously is still unclear. The relative contribution of these two mechanisms in driving feedback in the local Universe AGN is also unclear.}
  {We select a large and representative sample of 806 radio-AGN from the MaNGA survey, that provides integral field unit (IFU) optical spectra for nearby galaxies. We define radio-AGN as sources which have excess emission above that expected from star formation. We aim to study the feedback driven by radio-AGN on the galaxy's ionised gas, its location, and its relation to AGN properties. We also aim to disentangle the role of jets and radiation in these systems.} 
  {We use a sample of nearby radio-AGN from $L_\mathrm{1.4GHz}\approx10^{21}-10^{25}$\whz to trace the kinematics of the warm ionised gas phase using their \OIII emission line. We measure the \OIII line width and compare it to the stellar velocity dispersion to determine the presence and location of the disturbed gas. We investigate the dependence of radial profiles of these properties on the presence of jets and radiation, and their radio luminosities.}
  {We find most disturbed \OIII kinematics and proportion of disturbed sources up to a radial distance of 0.25\,\reff, when both radio and optical AGN are present in a source, and the radio luminosity is larger than $10^{23}$\whz. When either only radio or optical-AGN are present, the impact on \OIII is milder. Irrespective of the presence of an optical-AGN, we find significant differences in the feedback from high and low luminosity radio-AGN only up to a radial distance of 0.25\,\reff. }
  {The presence of more kinematically disturbed warm ionised gas in the central region of radio-AGN host galaxies, is related to both jets and radiation in these sources. We propose that in moderate radio luminosity AGN ($L_\mathrm{1.4GHz}\approx10^{23}-10^{25}$\whz) gas clouds pushed to high velocities by the jets (radiation) are driven to even higher velocities by the impact of radiation (jets) when both radio and optical-AGN are present. At lower luminosities ($L_\mathrm{1.4GHz}\approx10^{21}-10^{23}$\whz), the correlation between the disturbed ionised gas and enhanced radio emission could either be due to wind-driven shocks powering the radio emission, or low-power jets disturbing the gas.}


    \date{Received 5 December 2024 / Accepted 23 March 2025 }    

    \maketitle
\section{Introduction}
\label{introduction}

Feedback from active galactic nuclei (AGN) is a critical component of models of galaxy evolution. A fraction of the energy released during an AGN phase can couple with the gas of the host galaxy, preventing it from forming stars. This kind of negative feedback is used to explain the cessation of star formation in massive galaxies that would otherwise be too large and too massive (e.g. example \citealt{Silk1998,DiMatteo2005,Fabian2012}). The radiative and mechanical energy required for this feedback can be carried by accretion-driven winds, radiation pressure, collimated jets of plasma, etc (see \citealt{King2015, Harrison2024ObservationalInterpretation} for a review).\par

On intergalactic scales, the released energy can prevent the cooling and eventual accretion of gas from the hot halos surrounding galaxies, which is the fuel for star formation \citep{McNamara2012,McNamara2007}. On galactic scales, AGN-driven feedback can shock the gas clouds, making them turbulent and sometimes even pushing them to high velocities (outflows). This prevents the gas clouds from collapsing to form stars. These processes occur on different physical and time scales and are therefore crucial to understand how the AGN interact with their host galaxies. Over the past few years, many studies have found evidence for negative feedback on galactic scales, due to both radiatively efficient and inefficient AGN (e.g. \citealt{Venturi2021,Speranza2021,Morganti2021,Murthy2019}).\par

Statistical studies of large samples have found a positive correlation between the presence of AGN and its energy output, and the presence of kinematically disturbed gas (e.g. \citealt{Mullaney2013, Woo2017,Santoro2020,Kukreti2024}), albeit with spatially unresolved spectra. Although useful to understand the population as a whole, these unresolved spectra cannot provide information about the location of the disturbed gas or the affected stellar population in the galaxy. This spatial information is needed to understand the scales on which AGN-driven feedback occurs in galaxies. \par

Spatially resolved spectra from an integral field unit (IFU) can provide information about the gas and stellar emission at different locations in the host galaxy. This has led to the discovery of some trends between the spatial distribution of the disturbed gas and the energy released by the AGN. For example, a positive correlation between the size of the kinematically disturbed ionised gas region and the radiative luminosity has been found for AGN with radiatively efficient accretion (Seyferts, Quasars; e.g. \citealt{Kang2018,Kim2023UnravelingRelation,Gatto2024TheAGN}). For similar sources, positive correlations have also been found between the width of the emission line profiles tracing the disturbed gas and the radiative luminosity (e.g. \citealt{Wylezalek2020,Deconto-Machado2022IonisedCristino}). This illustrates the role of radiation in driving feedback in AGN host galaxies, however, such systematic spatially resolved studies in radiatively inefficient AGN, where jets are prominent, have been rarely conducted. It is important to study feedback in sources where both jets and radiation from the AGN are present to understand the role and relative contributions of these two mechanisms. \par

Studying feedback in radio-AGN hosts can help gain insight in this direction. In this paper, we define radio-AGN as sources with excess radio emission above that expected from star formation. However, we note that the definition is not uniform across the literature. Although the excess radio emission in radio-AGN is often due to collimated plasma jets launched during the active phase, in some low luminosity sources it can also originate from shock interactions between AGN-driven winds and the interstellar medium (ISM; \citealt{Zakamska2016}), relating the radio emission to the radiative output of the AGN. A large fraction of radio-AGN have radiatively inefficient accretion, but a significant proportion also show radiatively efficient accretion. This makes them ideal for understanding the role of both mechanical and radiative energy output of an AGN.\par

Many studies have found observational evidence for jet-driven feedback on the ionised gas in these galaxies using IFU data. However, these studies have mostly been focused on single objects or small samples (e.g. \citealt{Jarvis2019,Jarvis2021,Cresci2023BubblesXID2028,Girdhar2022QuasarDisc,Speranza2021, Nesvadba2021,Ruffa2022,Riffel2023,Ulivi2024Feedback0.15,Venturi2021}). Studies of large samples of radio-AGN with spatially resolved spectra have been hard to perform due to the lack of IFU data for a large sample of galaxies. This can now be achieved with the Mapping Nearby Galaxies at APO (MaNGA) survey (SDSS-IV), which is the largest IFU survey till date, covering a sample of $\sim$\,10,000 galaxies over a redshift range of $0.01<z<0.15$. This survey provides spatially resolved spectra from 3600$-$10300\AA, which covers emission lines like \OIII, \hbeta, \halpha, \Nii~etc, that are good indicators of AGN interaction and ionisation. Recently, \citet{Alban2024MappingCycleb} have identified AGN in MaNGA, using multi-wavelength data, and studied the warm ionised gas kinematics using the \OIII line. They have found evidence for enhanced ionised gas kinematics over galactic scales for radio-AGN compared to optical-AGN, suggesting that radio-AGN maybe tracing AGN at an advanced stage of the activity and feedback cycle.\par

We use the MaNGA survey data to study feedback from radio-AGN on the host galaxy's warm ionised gas ($T\,\sim\,10^{4}$\,K), using the \OIII$\mathrm{\lambdaup\lambdaup}$4958,5007\AA~doublet as a tracer. We use two radio surveys: the LOFAR Two-metre Sky Survey at 144\,MHz (LoTSS; \citealt{Shimwell2017,Shimwell2022}) and the Faint Images of the Radio Sky at Twenty-cm survey at 1400\,MHz (FIRST; \citealt{Becker1995TheCentimeters}), to identify the largest sample of radio-AGN from MaNGA. We trace the ionised gas using the \OIII emission line and determine the physical scales on which the radio-AGN impacts it. Our main aim is to disentangle the role of jets and radiation in driving feedback on warm ionised gas and determine the impact of these two mechanisms when they are present in the same source. The large size of these three surveys, combined with the high sensitivity of LoTSS to faint radio emission, allows us to cover a large range of radio luminosities from $L_\mathrm{1.4GHz}\approx10^{21}-10^{25}$\,\whz. This makes the sample a good representative of the radio-AGN population in the local Universe, and ideal for understanding their potential feedback. \par

This paper is structured in the following manner: Section 2 describes the sample construction and selection of radio-AGN, Section 3 describes the modelling of the stellar continuum and emission line profiles of the sources, Section 4 describes the diagnostics used to select the radio-AGN sample and Section 5 describes the radio properties of this radio-AGN sample. We then present the results for different source groups in Section 6 and discuss them in Section 7. Throughout the paper, we have used the $\Lambda$CDM cosmological model, with H$_\mathrm{0}$ = 70 km s$^\mathrm{-1}$ Mpc$^\mathrm{-1}$, $\Omega_\mathrm{M}=0.3$ and $\Omega_\mathrm{vac}=0.7$.

\section{Sample construction}
\label{sample construction}
This section outlines the approach used to select the radio-AGN, optical-AGN and control sample of non-AGN. MaNGA survey provides a wavelength coverage over 3600$-$10300\,\AA~with a spectral resolution of $R$\,$\sim$\,2000, corresponding to a spectral line-spread function width of about 70\kms~($1\sigma$). The IFU field of view diameter varies from $12\arcsec$ to $32\arcsec$. For more details on the survey, we refer the reader to \citet{Bundy2015OverviewObservatory}. Data reduction has been performed using the MaNGA Data Reduction Pipeline (DRP), described in \citet{Law2016THESURVEY,Law2021SDSS-IVAccuracy}. Next, spectral fitting is done by the MaNGA data-analysis pipeline (DAP) described in \citet{Westfall2019TheOverview}, which fits the stellar continuum and emission lines to the entire spectrum. Since this paper focuses on the \OIII gas kinematics, we use the emission line only spectra from the DAP, obtained after subtracting the stellar continuum from the reduced observed spectra. \citet{Sanchez2022SDSS-IVGalaxies} have constructed a catalogue of the host galaxy properties from a spectral analysis using the \texttt{pyPipe3D} pipeline \citep{Lacerda2022PyFIT3DPipeline}. We used properties such as stellar mass, effective radius, and ellipticities from their value-added catalogue. \par

Further, we remove duplicate observations from the catalogue. \citet{Alban2024MappingCycleb} have carried out a careful inspection of the MaNGA data and identified new duplicate observations, and combined them with the already known duplicates from \citet{Sanchez2022SDSS-IVGalaxies} and the survey website\footnote{\texttt{https://www.sdss4.org/dr17/manga/manga-caveats/}}. They have also removed sources flagged \texttt{CRITICAL} in the \texttt{MANGA\_DRP3QUAL} column of the DRP table. Finally, this gives us a sample of 9,777 sources.\par

To obtain statistically reliable results, selecting a large sample of radio-AGN from MaNGA is important. LoTSS has high sensitivity and source density, whereas FIRST has a larger area overlap with MaNGA. LoTSS and FIRST have 56\% and 97\% sky area overlap with MaNGA respectively. Therefore we use both surveys to maximise the number of radio cross-matches. We cross-matched the MaNGA sample with the LoTSS DR2 (6\arcsec angular resolution) and FIRST (5.4\arcsec angular resolution) catalogue using Topcat \citep{Taylor2005}, with a 6$\arcsec$ radius, which is equal to LoTSS and close to FIRST resolution. We cross-matched the sample down to flux densities of 0.3\,mJy in LoTSS and 0.5\,mJy in FIRST, to ensure that the detections have a $S/N>3$. This limit is chosen to maximise the number of cross-matches with MaNGA while keeping only reliable detections. At a search radius of 6$\arcsec$, we determine a contamination level from false-positive matches of $\sim$\,2.5\% for LoTSS cross-matches and 0.7\% for FIRST cross-matches \citep{Galvin2020CataloguingEra}. We found cross-matches for 2493 sources in LoTSS and 1155 sources in FIRST. LoTSS is sensitive to extended emission around sources and therefore provides a reliable measurement of the total flux density at 144\,MHz. However, FIRST has poor sensitivity to extended emission, meaning it could underestimate the total flux density at 1.4\,GHz. For more reliable estimates of the total flux density at 1.4\,GHz, we used the combined radio catalogue from \citet{Mingo2016}, which cross-matched the FIRST and NRAO VLA Sky Survey (NVSS; \citealt{Condon1998}). They use the flux densities from the lower resolution ($\sim$\,45\arcsec) NVSS survey, which recovers the total flux better than FIRST. Finally, we use the 144\,MHz and 1.4\,GHz total flux densities to estimate the $k$-corrected total radio luminosities, $L_\mathrm{144MHz}$ and $L_\mathrm{1.4GHz}$. \par

The diagnostics we use to select radio-AGN were constructed and optimised using the value-added spectroscopic catalogues produced by the group from the Max Planck Institute for
Astrophysics, and The Johns Hopkins University (MPA-JHU; \citealt{Brinchmann2004StellarDR2}). These include 
galaxy parameters measured using single fibre spectra from SDSS DR8. Since we use these diagnostics for radio-AGN selection, we use the MPA-JHU catalogue measurements for our galaxies, which include the $D_\mathrm{{n}}(4000)$, \OIII, \hbeta, \Nii~and \halpha~fluxes. Using a matching radius of 6$\arcsec$, we found cross-matches for 9072 sources in our sample. Since some MaNGA sources are missed by this cross-match, we also use 
the \citet{Sanchez2022SDSS-IVGalaxies} value-added catalogue which provides these measurements using a 2.5$\arcsec$ aperture, similar in size to the SDSS fibre. Although we do not use the \citet{Sanchez2022SDSS-IVGalaxies} catalogue measurements for all the sources, we note that using the 2.5$\arcsec$ aperture values for the entire sample only leads to a small fraction of sources being classified differently, but does not change our results.\par

In one of the diagnostics discussed in Section~\ref{selecting radio-AGN}, we use mid-infrared colours to separate star-forming and passive sources. For this, we use Wide-field Infrared Survey Explorer (WISE) data from the allWISE IPAC release (November 2013; \citealt{Cutri2021}). This provides magnitudes in three bands - W1 at 3.4$\mu$m, W2 at 4.6$\mu$m and W3 at 12$\mu$m, with an angular resolution of $6.1-6.5$\arcsec. To have reliable detections, we only selected sources with \texttt{cc\_flag=000} as suggested in the online user manual. We used a signal-to-noise ratio (S/N) threshold of 5 for W1 and W2, and 3 for W3 due to its poorer sensitivity. With these criteria, we obtained a WISE cross-match for 7616 sources.\par

Finally, we aim to construct radial profiles of \OIII line widths to study the spatial changes in \OIII kinematics. Galaxy morphology is an important property that can affect the gravitational motion of the gas, and therefore the spatial distribution of the \OIII line widths. This distribution can change depending on the relative strength of the bulge or disk component. To control for this, we added morphological classifications of the MaNGA galaxies from the Deep Learning DR17 Morphology catalogue \citep{DominguezSanchez2022SDSS-IVCatalogues,Fischer2019SDSS-IVMomentum}. We use the \texttt{T-Type} parameter for this purpose. Broadly, galaxies with \texttt{T-Type} $<0$ are ellipticals, \texttt{T-Type} $\sim0$ are S0s, and \texttt{T-Type} $>0$ are late-type. We control the \texttt{T-Type} distribution of the different groups studied in Section~\ref{results} to ensure that the effect of morphology is not driving the trends we observe. \par

\subsection{Selecting radio-AGN}
\label{selecting radio-AGN}
\begin{figure*}
  \centering
      \includegraphics[width=1.8\columnwidth]{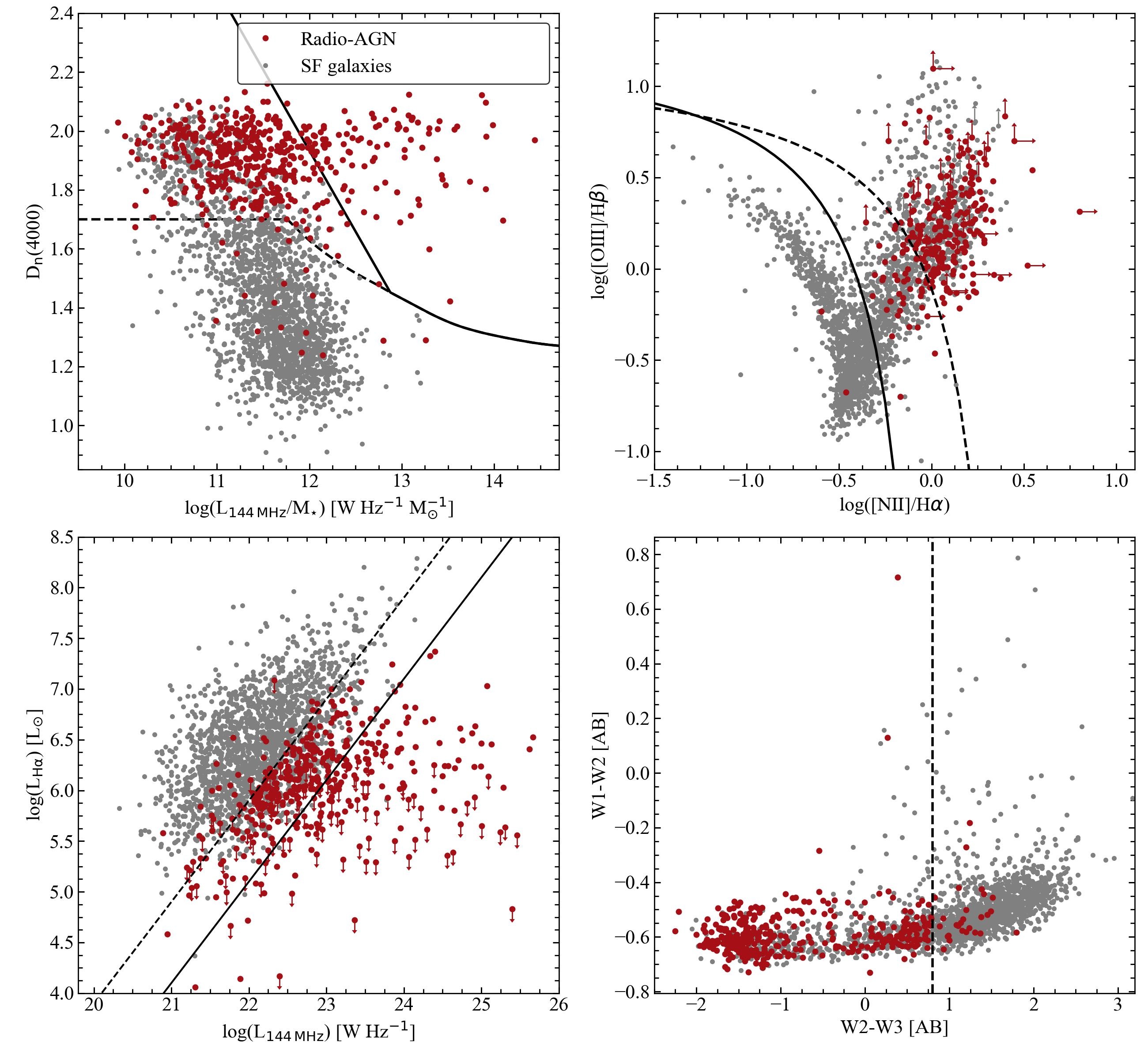}
      \caption{Diagnostic plots for selecting radio-AGN using LoTSS data. Red points mark the sources classified as radio-AGN after combining all four diagnostics, and grey points show the star-forming/radio-quiet AGN. (a) $D_\mathrm{{n}}(4000)$ vs $L_\mathrm{1.4GHz}$/$M_{\star}$ plot. The solid and dashed curves mark the radio-AGN, intermediate and SF/RQAGN division from \citet{Best2012} and \citet{Sabater2019}. (b) $L_{\textrm{H}\alpha}$ versus $L_\mathrm{1.4GHz}$ plot, with the separation lines from \citet{Sabater2019}. (c) BPT diagram where the solid curve shows the semi-empirical relation from \citet{Kauffmann2003}, and the dashed curve shows the maximum starburst curve from \citet{Kewley2006}. (d) WISE colour-colour plot for sources with the division line from \citet{Sabater2019}.}
    \label{diagnostic_lotss}
  \end{figure*}

In the local Universe, star-forming galaxies (SFGs) dominate a radio-selected sample up to $L_\mathrm{1.4GHz}$\,$\approx$\,$10^{23}$ \whz \citep{Sadler2002,Best2005}. Most MaNGA sources with a radio cross-match have a luminosity lower than this value. Therefore, we first need to select a clean sample of radio-AGN, i.e. sources where radio emission exceeds that expected from star formation and can be attributed to the AGN in the galaxy. To do this, we use a combination of four diagnostics created by \citet{Best2005,Best2012} and further developed by \citet{Sabater2019}. We point the reader to these references for a detailed discussion of these methods. We first select radio-AGN from the LoTSS cross-matched sample using diagnostics shown in Fig.~\ref{diagnostic_lotss}, and then from the FIRST cross-matched sample using diagnostics shown in Fig.~\ref{diagnostic_first}. \par

The first diagnostic is the `$D_\mathrm{{n}}(4000)$ versus $L_\mathrm{radio}$/$M_{\star}$' method. Here, $D_\mathrm{{n}}(4000)$ is the spectral break at 4000\,\AA, which is an indicator of the mean stellar age. $L_\mathrm{radio}$/$M_{\star}$ is the ratio of radio luminosity and stellar mass. These quantities depend on the star formation rate of a galaxy and can be used to locate SFGs on a plane. \citet{Best2005} showed that SFGs broadly occupy the same region on a plot of these two quantities for a large variety of star formation histories. But in the case of radio-AGN, excess radio emission from the AGN would increase the $L_\mathrm{radio}$/$M_{\star}$ value, separating them from the SFGs. Based on this idea, \citet{Best2005} and \citet{Kauffmann2008} obtained a track to separate SFGs from radio-AGN. \citet{Sabater2019} selected radio-AGN from LoTSS-DR1, and added another track that follows the original track till $D_\mathrm{{n}}(4000)=1.7$, then continues horizontally. They helped maximise agreement with the more sophisticated selection of SFGs by \citet{Gurkan2018} for the H-ATLAS sample and select low-luminosity radio-AGN that have high $D_\mathrm{{n}}(4000)$ values. Since MaNGA sources typically have low radio luminosities, we also used the track of \citet{Sabater2019}\footnote{Both tracks were provided by Philip Best (private communication)}. For $L_\mathrm{radio}$, we used the 144\,MHz and 1.4\,GHz luminosity respectively, for LoTSS and FIRST cross-matched sources. For FIRST cross-matched sources, we scaled the diagnostic tracks to 1.4\,GHz assuming an optically thin spectrum modelled as $S\propto\nu^{\alpha}$, with a spectral index of $\alpha=-0.7$. Varying $\alpha$ between -0.5 and -1 does the change the classification of a few sources but does not affect our results. In the `$D_\mathrm{{n}}(4000)$ versus $L_\mathrm{radio}$/$M_{\star}$' plot in Fig.~\ref{diagnostic_lotss} and \ref{diagnostic_first}, sources above the solid line are classified as radio-AGN, sources below the dashed line are classified as SFGs, and sources between the two lines are classified as intermediate. \par

The second diagnostic is the $L_{\textrm{H}\alpha}$ versus $L_\mathrm{radio}$ plot. The star formation rate of massive stars can determine both \halpha~and radio luminosities of galaxies. Therefore sources with excess radio emission contribution from an AGN can be separated on a plot of these two properties. We again use the relations from \citet{Sabater2019} for the 144\,MHz luminosity of LoTSS cross-matched sources: log($L_{\textrm{H}\alpha}$/L$_{\odot}$) = log($L_\mathrm{144MHz}/$\whz$) - 16.1$ and log($L_{\textrm{H}\alpha}$/L$_{\odot}$) = log($L_\mathrm{144MHz}/$\whz$) - 16.9$. These relations were derived to again maximise the agreement with the results of \citet{Gurkan2018} for the H-ATLAS sample, to avoid misclassification of low luminosity sources. In the plots shown in Fig.~\ref{diagnostic_lotss} and \ref{diagnostic_first}, sources with less $L_{\textrm{H}\alpha}$ than the lower line are classified as radio-AGN, sources between the two lines as intermediate and above the upper line are classified as star-forming. We used \halpha~non-detections in cases where they provided useful upper limits on $L_{\textrm{H}\alpha}$. For FIRST cross-matched sources, we scaled the relations mentioned above to 1.4\,GHz and performed the same classification. \par

The third diagnostic used is the BPT diagram, where we use the \Nii/\halpha~and \oiiihbeta~emission line ratios to separate sources ionised by AGN or star-formation. We used the division from \citet{Kauffmann2003} and maximum starburst relation form \citet{Kewley2006}. Sources with larger ratios than the maximum starburst curve were classified as AGN, between the two curves as intermediate and below the lower curve as star-forming. Although this does not use any radio property, it can still give information about the presence of AGN or star formation in the source. We again used non-detections in cases where they provided useful upper limits. \par

\begin{table}[!ht]
\centering
\caption{Diagnostic combinations for LoTSS sources}
         \label{diagnosticcomb_lotss}
\renewcommand{\arraystretch}{1.15}
\setlength{\tabcolsep}{4pt}
\begin{tabular}{cccccc}
    \hlineB{3}
    \noalign{\vspace{0.05cm}}
    \hline
    \noalign{\smallskip}
     $D_\mathrm{{n}}(4000)$  & BPT  & $L_{\textrm{H}\alpha}$  & WISE & Number &  Final \\
     vs $L_\mathrm{144MHz}$/$M_{\star}$ &   & vs $L_\mathrm{144MHz}$ & col-col &  & class\\
     
    \noalign{\smallskip}
    \hlineB{3}

    \noalign{\smallskip}
    \noalign{\smallskip}

AGN & AGN & AGN & AGN & 14 & AGN\\ 
AGN & Uncl. & AGN & AGN & 13 & AGN\\ 
AGN & Uncl. & AGN & Uncl. & 10 & AGN\\ 
AGN & Uncl. & Uncl. & AGN & 20 & AGN\\ 
AGN & Uncl. & Uncl. & Uncl. & 11 & AGN\\ 
Int. & AGN & AGN & SF & 13 & AGN\\ 
Int. & AGN & Int & AGN & 116 & AGN\\
Int. & AGN & Int. & SF & 28 & SF\\ 
Int. & AGN & Int. & Uncl. & 12 & AGN\\  
Int. & AGN & SF & AGN & 170 & SF\\ 
Int. & Int. & Int. & AGN & 31 & AGN\\
Int. & Int. & Int & SF & 28 & SF\\
Int. & Int. & SF & AGN & 64 & SF\\ 
Int. & Uncl. & Int. & AGN & 29 & AGN\\ 
Int. & Uncl. & SF & AGN & 21 & SF\\ 
Int. & Uncl. & Uncl. & AGN & 80 & AGN\\
Int. & Uncl. & Uncl. & Uncl. & 15 & AGN\\ 
SF & AGN & SF & AGN & 40 & SF\\ 
SF & AGN & Int. & AGN & 12 & AGN\\ 
SF & AGN & Int. & SF & 17 & SF\\ 
SF & Int. & SF & Uncl. & 18 & SF\\ 
SF & AGN & SF & SF & 43 & SF\\ 
SF & Int. & Int. & SF & 65 & SF\\ 
SF & Int. & SF & AGN & 55 & SF\\ 
SF & Int. & SF & SF & 142 & SF\\ 
SF & SF & Int. & SF & 124 & SF\\ 
SF & SF & SF & AGN & 43 & SF\\ 
SF & SF & SF & SF & 829 & SF\\ 
SF & SF & SF & Uncl. & 54 & SF\\ 
SF & Uncl. & Int & SF & 17 & SF\\ 
Uncl. & Uncl. & Uncl. & Uncl. & 27 & Uncl.\\ 

    \noalign{\smallskip}
    \noalign{\smallskip}
   
    \hline
    \noalign{\vspace{0.05cm}}  
    \hlineB{3}
    \end{tabular}
    \flushleft
    Note. Combinations for classification of a source using the diagnostic diagrams discussed in Section~\ref{selecting radio-AGN}, and the final classification assigned. Only groups with more than 40 sources are shown here. The same table for FIRST detected sources is shown in Table~\ref{diagnosticcomb_first}     
\end{table}

\begin{table}[!ht]
\centering
\caption{Classification of LoTSS cross-matched sources}
         \label{diagnostictable_lotss}
\renewcommand{\arraystretch}{1.15}
\setlength{\tabcolsep}{4pt}
\begin{tabular}{ccccc}
    \hlineB{3}
    \noalign{\vspace{0.05cm}}
    \hline
    \noalign{\smallskip}
     Diagnostic & AGN & Int. & SF & Uncl.\\
     & \multicolumn{4}{c} {(No. of overall radio-AGN)} \\ 
     
    \noalign{\smallskip}
    \hlineB{3}

    \noalign{\smallskip}
    \noalign{\smallskip}
  
  $D_\mathrm{{n}}(4000)$ vs $L_\mathrm{144MHz}$/$M_{\star}$ & 102 & 705 & 1675 & 11  \\ 
   & (102) & (378) & (19) & -  \\ 
    \noalign{\smallskip}
  $L_{\textrm{H}\alpha}$ vs $L_\mathrm{144MHz}$ & 148 & 623 & 1583 & 139 \\
   & (134) & (216) & (24) & (125)  \\ 
    \noalign{\smallskip}
  BPT & 593 & 484 & 1122 & 294 \\
   & (220) & (52) & (6) & (221) \\
       \noalign{\smallskip}
  WISE col-col & 839 & - & 1433 & 221 \\
   & (374) & - & (27) & (98)  \\ 
    \noalign{\smallskip}
    \noalign{\smallskip}
   
    \hline
    \noalign{\vspace{0.05cm}}  
    \hlineB{3}
    \end{tabular}
    \flushleft
    Note. Number of sources classified by each diagnostic discussed in Section~\ref{selecting radio-AGN}. The different classes are - AGN, intermediate (Int.), star-forming (SF) and unclassified (Uncl.). The numbers in brackets are sources from each group classified as radio-AGN after combining the four diagnostics.     
\end{table}
\begin{figure}
\centering
  \includegraphics[width=\columnwidth]{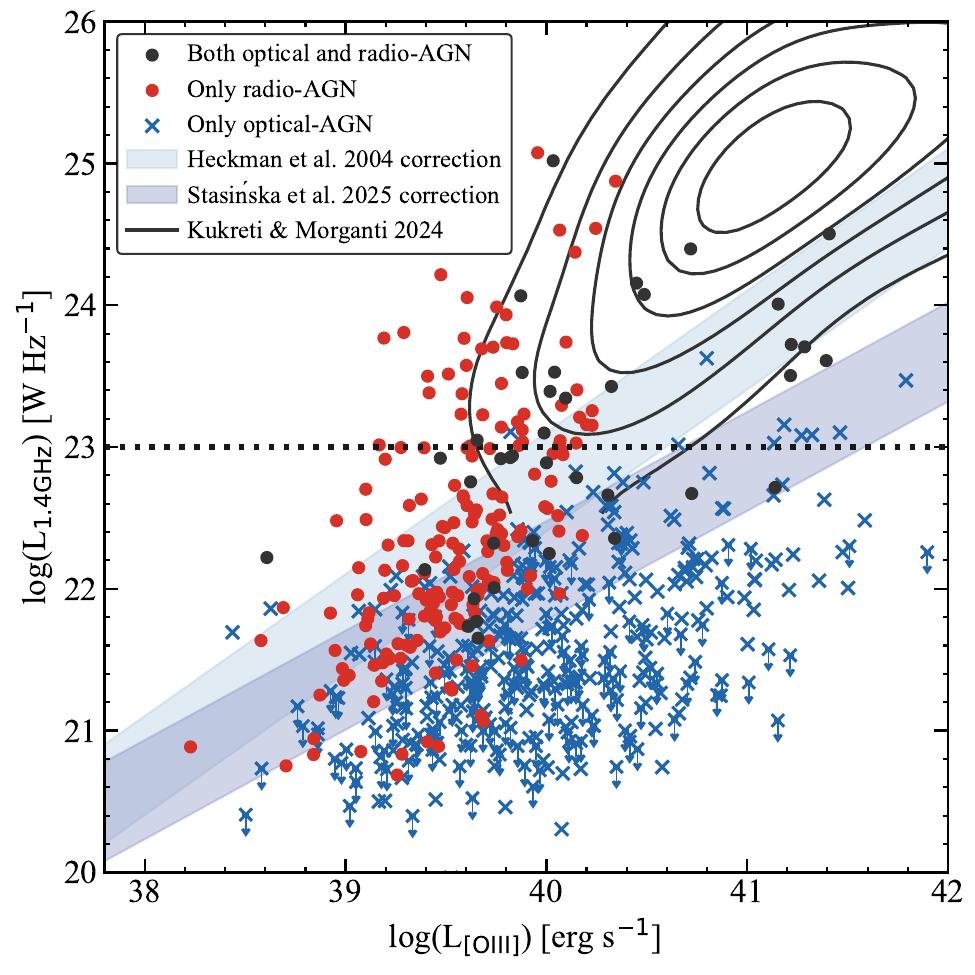}
  \caption{1.4\,GHz and \OIII luminosities of the sources with an \OIII detection. Source with radio+optical AGN, radio-AGN and optical-AGN are shown. For optical-AGN without a radio detection, the arrows show the upper limits on $L_\mathrm{1.4GHz}$. The shaded regions show the expected radio luminosity from the shocks scenario using the \cite{Nims2015ObservationalNuclei} model, where the lower and upper edges show the limits for 5\% and 25\% coupling efficiency between AGN luminosity and wind kinetic luminosity, respectively. Different regions show the \OIII to bolometric luminosity conversion factors from \cite{Heckman2004} and \cite{Stasinska2025OpticallySDSS}. The horizontal dotted lines show the luminosity divisions used in Section~\ref{results}. To highlight the low-to-moderate luminosity nature of the sources in our study, we also show the parameter space occupied by the radio-AGN sample from \citet{Kukreti2024} in black contours, which studied the relation between radio properties and feedback on \OIII out to $z=0.8$.}
  \label{source prop}    
\end{figure}

The final diagnostic is the WISE colour-colour diagram, between the $W1-W2$ and $W2-W3$ colours. Radio-AGN host galaxies are typically ellipticals, with low levels of star formation. These can be separated from star-forming galaxies based on their $W2-W3$ colour \citep{Yan2013}. \citet{Sabater2019} modified the division of \citet{Herpich2016} based on comparison with the results of \citet{Gurkan2018} for the H-ATLAS field as mentioned before. We use their division of $W2-W3$ (AB) = 0.8.\par

We classify the LoTSS and FIRST cross-matched sources independently. This is done to maximise the number of radio-AGN selected from the sample. In each diagnostic (except the WISE diagram), a source can be labelled as radio-AGN, intermediate, star-forming or unclassified. This gives a total of 192 possible combinations. These combinations are used to give a final classification, following the approach of \citet{Sabater2019} and \citet{Best2012}. This approach gives the most weight in classification to the `$D_\mathrm{{n}}(4000)$ vs $L_\mathrm{radio}$/$M_{\star}$' and `$L_{\textrm{H}\alpha}$ vs $L_\mathrm{radio}$' diagnostic. Sources classified by either of these plots as an AGN are classified as a radio-AGN in the final classification. The BPT and WISE colour-colour diagnostics have the least weight (as they have no radio information) and are used only when the first two diagnostics give inconclusive classifications. Their main purpose at this stage is to check whether a source has SF or not. For example, if both of the first two radio diagnostics classify a source as ‘Intermediate’, we use the BPT and WISE colour-colour plot to check if the source lies in the SF region or not, to determine if the radio emission can be explained by SF. It is worth noting that the main aim of this selection technique is to select a clean sample of radio-AGN and is not necessarily complete as it might miss some radio-AGN that reside in SF galaxies.\par 

The different combinations of the diagnostics, the number of sources in each group and the final classification are summarised in Table~\ref{diagnosticcomb_lotss} for LoTSS sources and Table~\ref{diagnosticcomb_first} for FIRST sources. The number of sources classified in each diagnostic, and the final number of radio-AGN from LoTSS and FIRST cross-matches, are summarised in Table~\ref{diagnostictable_lotss} and ~\ref{diagnostictable_first} respectively. Finally, we select 499 LoTSS and 512 FIRST sources as radio-AGN, which leads to a combined sample of 806 radio-AGN identified either with LoTSS or FIRST. Of these, 205 are common in both LoTSS and FIRST samples, 294 are unique in LoTSS and 307 are unique in the FIRST radio-AGN sample. For radio-AGN with only a LoTSS detection, we estimate the $k$-corrected 1.4\,GHz luminosity by extrapolating the flux from 144\,MHz using a power law of the form $F_{\nu}\propto\nu^{\alpha}$ with a spectral index $\alpha=-0.7$. This allows us to assess them together with radio-AGN that have FIRST detections. The distribution of this radio-AGN sample on the $L_\mathrm{1.4GHz}$ vs $L_{\textrm{\OIII}}$ plot is shown in Fig.~\ref{source prop}. \par

Our radio-AGN sample mostly covers a $L_\mathrm{1.4GHz}$ range of $\approx10^{21}-10^{24}$\,\whz, with the majority of sources below $10^{23}$\,\whz, which can be classified as low-luminosity radio-AGN. Over the past years, studies of feedback in radio-AGN host galaxies with IFU data have mostly focused on moderate to high luminosity ($\gtrsim10^{25}$\,\whz) radio-AGN, as discussed in the Introduction section. However, with the combination of large IFU surveys like MaNGA and sensitive radio surveys like LoTSS, it is now possible to perform such studies for low-to-moderate luminosity radio-AGN, which make up the majority of the radio-AGN population in the local Universe \citep{Best2005,Sadler2002,Sabater2019}. Therefore, our study of mostly low-luminosity radio-AGN can help us understand the role of this population in driving feedback. The low luminosity nature of our sample is highlighted in Fig.~\ref{source prop} where we also show the sample from \citet{Kukreti2024} for comparison. Radio-AGN in this sample were selected using the same diagnostics to study feedback on \OIII up to $z=0.8$ using single fibre SDSS spectra. \par

Although the radio emission in radio-loud AGN is attributed to jets of relativistic plasma, the origin is less clear for radio-quiet AGN. One interesting explanation for radio emission in these sources is the shock scenario. This includes radiatively accelerated winds driving shocks into the host galaxy medium, which accelerate particles that then produce synchrotron emission \citep{Faucher-Giguere2012TheNuclei,Zakamska2014,Zakamska2016,Zubovas2012ClearingGalaxy}. This scenario can explain the origin of radio-emission in low-to-moderate radio luminosity sources, similar to the ones in our sample. In Fig.~\ref{source prop}, we also show the expected shock-generated radio luminosities from the fiducial models of \citet{Nims2015ObservationalNuclei}. This model estimates the expected radio luminosity for a certain AGN bolometric luminosity. However, there are significant uncertainties in the coupling efficiencies between the AGN bolometric luminosity, wind kinetic luminosity and energy of the accelerated electrons. The conversions determining the bolometric luminosity from the observed \OIII luminosity are also uncertain. We show the \cite{Nims2015ObservationalNuclei} model for two different conversions of \OIII to bolometric luminosity, from \cite{Heckman2004} and \cite{Stasinska2025OpticallySDSS}, and covering a range of coupling efficiencies. We find that $\sim$40\% of the sources with radio+optical AGN (section~\ref{optical agn sample}) lie beyond the radio luminosity limit of this model, and very likely have radio emission dominated by jets. In the rest, the radio emission could also be due to shocks. We discuss this further in Section~\ref{discussion}.\par

\subsection{Other MaNGA radio-AGN catalogues}
\label{other manga radio-AGN}

Multiple studies in the literature have selected radio-AGN samples from the MaNGA catalogue, and here we compare those samples to ours. \citet{Comerford2024AnSUB/SUB} cross-matched the MaNGA catalogue with the radio-AGN catalogue of \citet{Best2012} which itself is selected using SDSS DR7 source catalogue with the FIRST and NVSS surveys. They find 221 of these radio-AGN in the MaNGA catalogue. We cross-matched the \citet{Best2012} catalogue with our 9,777 MaNGA sources using a 6\arcsec match radius and found 214 radio-AGN instead. The discrepancy could be due to the different number of MaNGA sources used to start with. Out of these 214 sources, 199 are selected as radio-AGN by our selection criteria with FIRST data. Of the remaining 15 sources, 13 do not have a cross-match in FIRST in our sample, likely due to a difference in the cross-matching criteria from \citet{Best2012}. The remaining 2 sources have a FIRST crossmatch but are not classified as radio-AGN in our sample. Although our selection criteria are based on \citet{Best2012}, we use the modified selection criteria by \citet{Sabater2019}, which attempt to include more low luminosity radio-AGN misclassified as SF before. We also select sources below the 5\,mJy threshold used by \cite{Best2012}, which would explain the larger number of radio-AGN we identified compared to \citet{Comerford2024AnSUB/SUB}. Broadly, almost all the radio-AGN in the \citet{Comerford2024AnSUB/SUB} sample are included in ours. \par
A sample of 307 radio-AGN has also been selected from MaNGA by \citet{Mulcahey2022StarMaNGA} using the same diagnostics as we use and LoTSS DR2 data. Compared to the radio-AGN selected using LoTSS data in our sample, we find an overlap of 189 sources. The sources that do not overlap are low luminosity ($L_\mathrm{144MHz}\approx10^{21}-10^{23}$\,\whz) sources in our sample. We found that the radio luminosities in that study have been overestimated which could affect the classification and explain this discrepancy. However, this does not affect the results of that study \footnote{private communication.}. \par

 \begin{figure*}
    \centering
    \includegraphics[width=2\columnwidth]{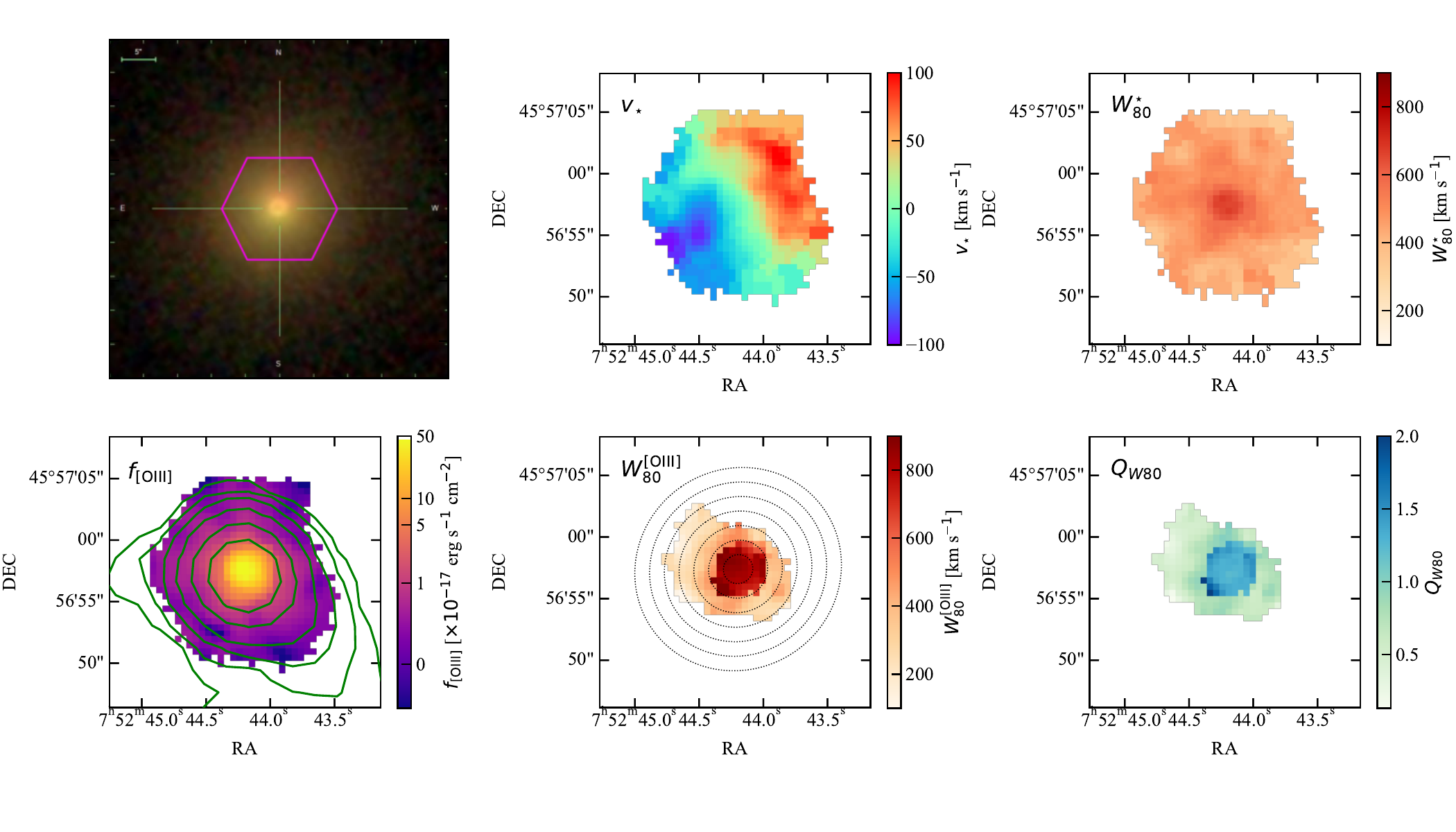}
    \caption{Maps of a radio-AGN from the sample (plateifu = 8714-3704). The top left image shows the SDSS optical colour image of the host galaxy, with the MaNGA FoV marked with magenta. The other panels show the stellar and \OIII emission line properties. Only spaxels with a \OIII detection of $S/N>5$ are shown and used throughout the paper for analysis. The green contours show the radio emission from FIRST at 1.4\,GHz in the bottom left panel. The elliptical annuli regions used for constructing radial profiles are shown in the bottom middle panel. The same plot for a non-AGN source is shown in Fig.~\ref{oiiimaps_nonagn} to illustrate the differences in their gas kinematics.}
    \label{oiiimaps}
 \end{figure*}

Recently, \citet{Alban2024MappingCycleb} have also selected a sample of 288 radio-AGN from the MaNGA catalogue with a \OIII $S/N>7$, by comparing the star formation rates (SFRs) estimated from 1.4\,GHz (FIRST+NVSS) and \halpha~luminosities (obtained from \citet{Sanchez2022SDSS-IVGalaxies} catalogue). On a plot of the two SFRs, they select sources with a radio SFR more than 0.5\,dex away from the 1:1 relation. Out of their 288 sources, 97 are also classified as radio-AGN in our sample using FIRST, and 159 are classified as SF/radio-quiet. This discrepancy is due to the significantly different selection methods and the different FIRST detection threshold they use of 1\,mJy. Indeed, most of the disagreeing sources in the two samples lie close to the 0.5\,dex division line on the SFR plot of \citet{Alban2024MappingCycleb}. Thirty sources in their sample do not have a FIRST crossmatch in our sample. This discrepancy is because they also cross-match with the NVSS catalogue, which gives them more sources with a radio detection at 1.4\,GHz. We compare and discuss the \citet{Alban2024MappingCycleb} catalogue in more detail in Section~\ref{discussion}, since they also measure the radial profiles of the \OIII line width. Recently, \citet{Suresh2024RadioRatesb} have also selected a sample of radio-AGN using a similar approach of comparing radio and \halpha~based SFRs, but used a 1\,dex threshold from the 1:1 line. 

\subsection{Optical-AGN sample}
\label{optical agn sample}
Since we aim to understand the complementary roles of jets and radiation in driving feedback, we also construct a sample of optical-AGN using the BPT classification in Section~\ref{selecting radio-AGN}. We first identify all sources with an AGN classification in the BPT diagram. We then select only sources with an \halpha~equivalent width greater than 3\AA. This is done to ensure that the ionisation is due to an AGN in these sources, as \citet{CidFernandes2010, CidFernandes2011AAGN} have shown that sources with weaker \halpha~emission that lie in the AGN (LINER) region of the BPT diagram, can have ionisation from hot low mass evolved stars, instead of an AGN. This gives us a sample of 482 optical-AGN, out of which 73 are also classified as radio-AGN. We therefore have 409 sources that are optical-AGN. We note that an optical-AGN sample using MaNGA data has also been selected by \citet{Alban2023ClassifyingApertures} using different aperture sizes. We find that 369 of our optical-AGN sources are also classified as AGN in their study. However, 113 of our sources are not classified as an AGN in their selection. These sources have low signal-to-noise line detections and \OIII luminosities. Therefore, this difference is likely a result of the different data used for selection. They use a 2\,kpc aperture in the central region to classify sources, whereas we use the data from the MPA-JHU catalogue, which is based on single fibre SDSS data. Since the single fibre data provides integrated fluxes over larger areas ($\sim2-8$\,kpc) at these redshifts, it can recover lower flux emission with higher signal-to-noise than IFU data. \par

Finally, we also split the radio-AGN sample into those with and without an optical-AGN, using the same criteria as above. This helps us disentangle the role of jets and radiation in Section~\ref{results}. Out of the 806 radio-AGN selected above, 73 are also optical-AGN. For the rest of the paper, we refer to sources with both radio and optical-AGN as radio+optical AGN. Sources with either only a radio or optical-AGN are simply referred to as radio or optical-AGN. \par

The selection methods used above can miss radio+optical AGN that have low radio luminosities but are optically bright. The extra contribution from AGN to the \halpha~luminosity, on top of any star-formation, might shift the sources vertically upwards on the $L_{\textrm{H}\alpha}$ versus $L_\mathrm{radio}$ plot, into the SF region. Therefore, our selection techniques would not be able to identify the radio-AGN in these sources. This has been observed before for quasars with low radio luminosities by \citet{Jarvis2021}, and would require high-resolution (sub-arcsecond) radio imaging to confirm the presence (or absence) of a radio-AGN. We note that such missed sources could contaminate our optical-AGN sample. However, the selection techniques used still allow us to select a clean sample of radio and radio+optical AGN. \par

The presence of broad emission lines from the type 1 AGN in the sample can affect the derived measurements of the host galaxy properties. We test for any systematic bias introduced by the presence of broad line AGN in the $D_\mathrm{{n}}(4000)$, stellar velocity dispersion and stellar mass measurements by comparing the broad line optical (and radio+optical) AGN with the narrow line AGN. We use the broad-line galaxy catalogue of MaNGA from \citet{Fu2023AAGNs}, which contains 135 galaxies. Comparing the broad-line and narrow-line AGN, we find no systematic bias in the measurements mentioned above and conclude that this does not contaminate our radio-AGN sample selection.

\subsection{Non-AGN control sample}
\label{controlsample}

To test whether any disturbed \OIII is associated with the presence of a radio-AGN, we also select a control sample of non-AGN galaxies for comparison. These sources don't necessarily have radio detection. We first only select sources classified as SF on the BPT diagram. We remove all the sources classified as radio-AGN in Section~\ref{selecting radio-AGN}. Then, we remove the AGN selected using mid-infrared data from WISE (123) and X-ray data from BAT (29), compiled by \citet{Comerford2024AnSUB/SUB}. Lastly, we also remove any broad-line galaxies (135) present in the catalogue of \citet{Fu2023AAGNs}, since these are mostly AGN as well. After visual inspection of the optical images and the \OIII maps shown in Fig.~\ref{oiiimaps} and \ref{oiiimaps_nonagn}, we also removed any mergers where the \OIII kinematics was affected by the interactions. We then restrict the sample to $M_{\star}>10^{11}$\,M$_{\odot}$ and $\sigma_{\star}>150$\,\kms, to match the distributions of our radio-AGN sample. This gives a total of 63 non-AGN sources. We estimate that the radio non-detections in the non-AGN sample have a radio luminosity value below $L_\mathrm{1.4GHz}=3\times10^{22}$\whz. An example of a non-AGN source is shown in Fig.~\ref{oiiimaps_nonagn}.

\section{\OIII spectra modelling and analysis}
\label{oiii modelling}

We now characterise the \OIII spectra of the MaNGA galaxies. The presence of disturbed gas kinematics is often evident in broad low amplitude components of the emission line. This requires careful modelling of the \OIII line profile. Although the DAP catalogue provides measurements for emission lines, they use only a single component while modelling the line profiles, which is insufficient for our purpose. Therefore, we use the best-fit models from \citet{Alban2024MappingCycleb}, who fit the \OIII spectrum of each spaxel with up to two Gaussian components. They use the least-squares algorithm to fit the components to both lines in the \OIII doublet simultaneously. We refer the reader to Section 3 of their paper for more details on the fitting procedure. \par

After determining the best-fit model, we estimate the $W_{80}$ widths of the \OIII profiles. This line width encloses 80\% of the total flux and is defined as the difference between the velocities that enclose 10\% and 90\% of the cumulative flux, i.e. $W_{80} = v_{90}-v_{10}$. Using $W_{80}$ allows us to include emission from broad components that are likely tracing kinematically disturbed gas, while also not being too sensitive to the low $S/N$ emission. It also enables comparison between profiles fit with different numbers of components. For a single Gaussian component, $W_{80}$ is related to the velocity dispersion $\sigma$ as $W_{80} = 2.563\times\sigma$, but the relation is not straightforward for multiple components. Using the approach outlined in \citet{Sun2017}, we correct the \width~values using the instrumental spectral resolution of $\approx$\,70\kms and subtract the equivalent \width~value in quadrature from the \OIII value. An example of a corrected \oiiiwidth~map for a source in our radio-AGN sample is shown in Fig.~\ref{oiiimaps}. \par

\begin{figure*}
    \centering
    \begin{subfigure}{1.7\columnwidth}
      \includegraphics[width=\columnwidth]{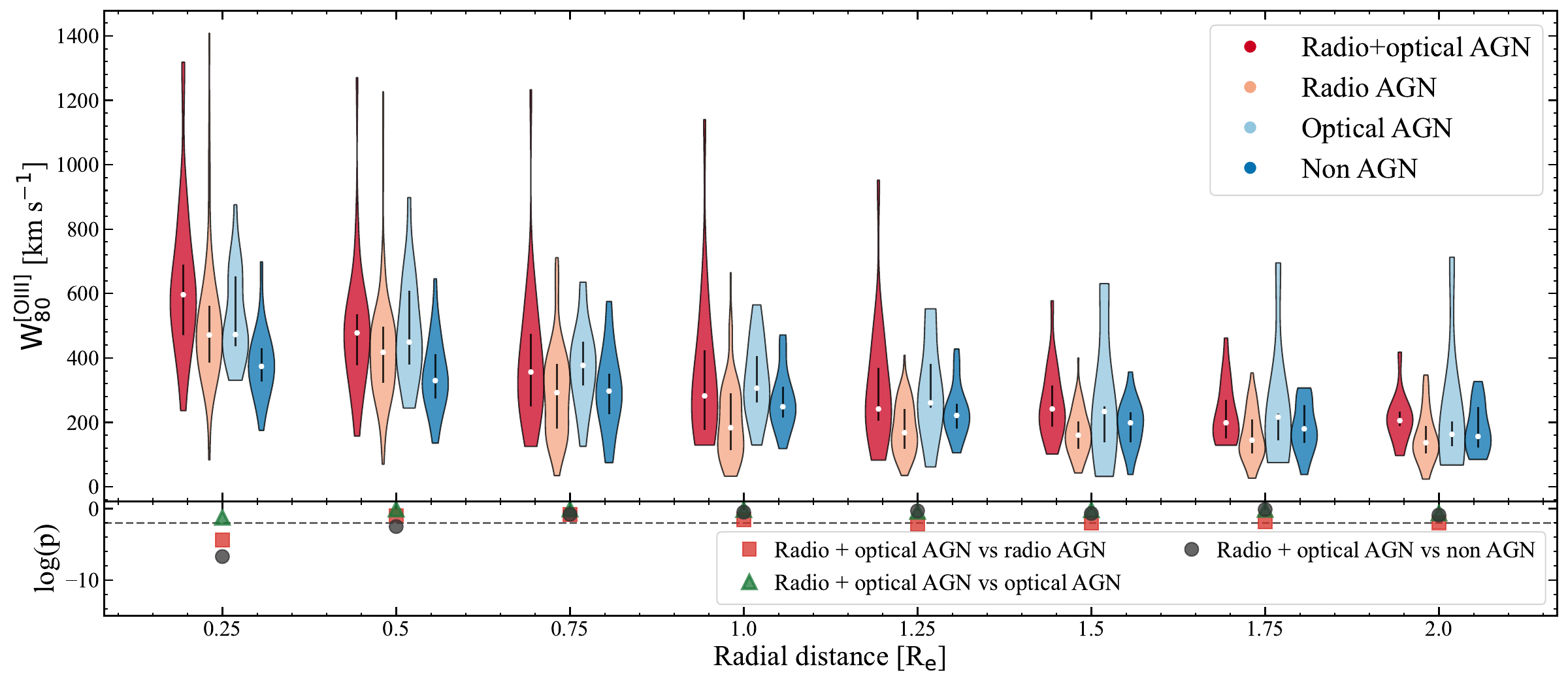}
      \caption{Radial profiles of \oiiiwidth }
      \label{ragn_radial_oiiiw80}
    \end{subfigure}
    \begin{subfigure}{1.7\columnwidth}
      \includegraphics[width=\columnwidth]{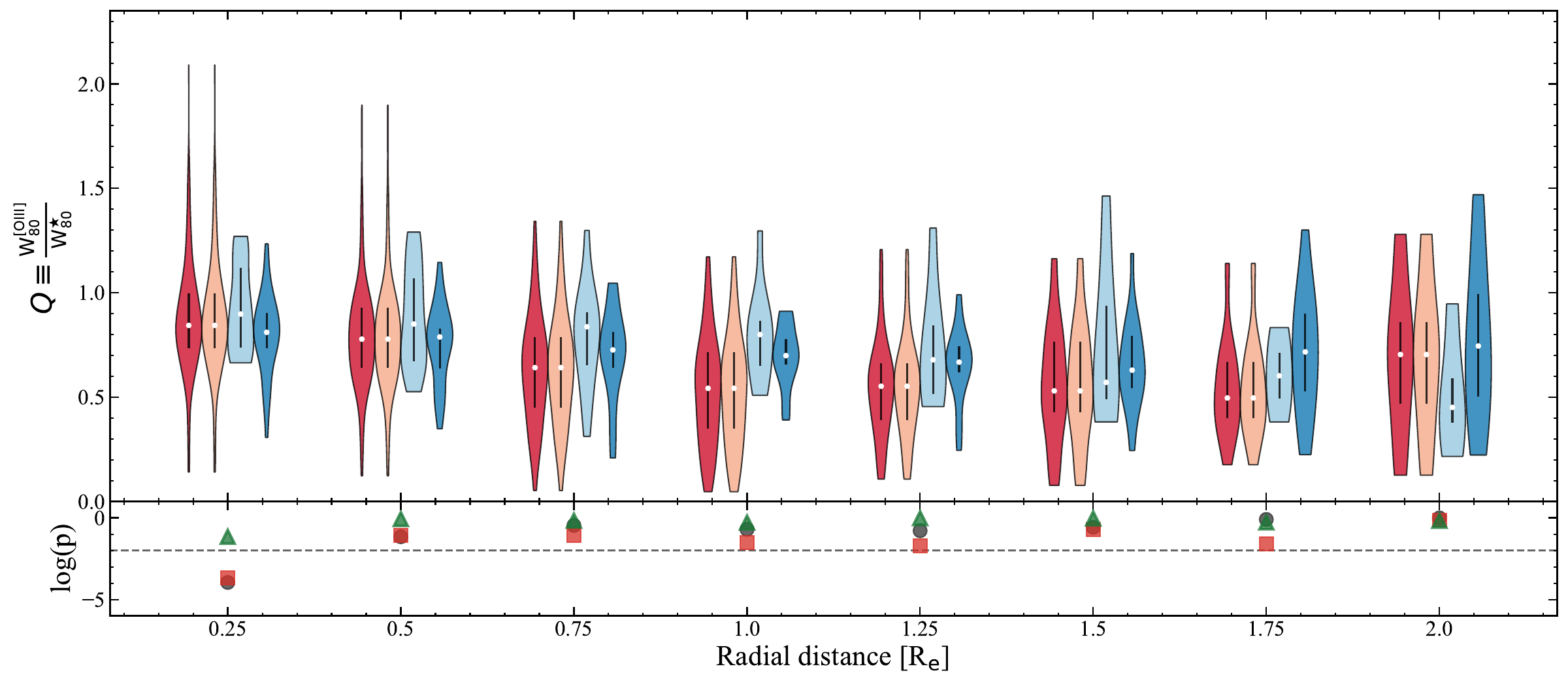}
      \caption{Radial profiles of \q }
      \label{ragn_radial_q}
    \end{subfigure}

    \begin{subfigure}{1\columnwidth}
      \includegraphics[width=\columnwidth]{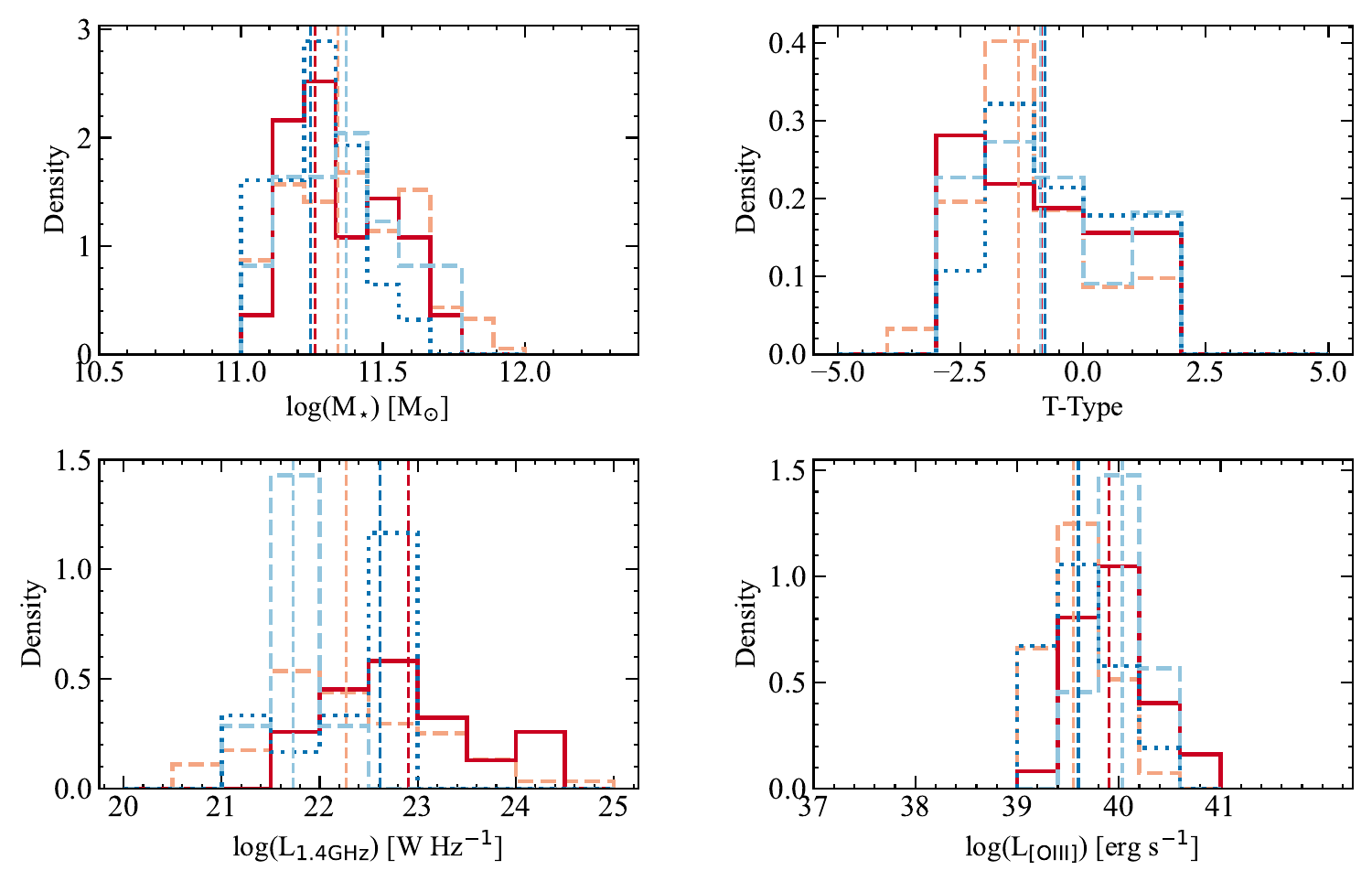}
      \caption{Host galaxy properties}
      \label{ragn_prop}
    \end{subfigure}
    \caption{Comparison of the four groups with: optical and radio-AGN, radio-AGN, optical-AGN and non-AGN. Radial profiles of \oiiiwidth ~(a) and \q~(b) values, in steps of 0.25 \reff. The violin plots show the distributions for every group at every radial point, the white circle shows the median values and the black lines show the range from the 25th to 75th percentile of the distributions. The bottom panel show the p-values for a 2-sample KS test at every radial point, with the two sample combinations marked in the legend. The horizontal dashed line marks the 99\% confidence level p-value. (c) Host galaxy properties of the source groups, showing the stellar mass, host galaxy T-type, radio luminosity and \OIII luminosity. Vertical lines show the median values. The colours represent the same sources as in the top panels. All source luminosities are only for radio and \OIII detections. }
    \label{radialprof_ragn}
 \end{figure*}

Next, to inspect spatial changes in the line widths of \OIII, we construct radial profiles of $W_{80}$ using the procedure described in \citet{Alban2024MappingCycleb, Alban2023ClassifyingApertures}. These profiles are constructed using elliptical annuli apertures of widths equal to $0.25\times R_\textrm{{eff}}$, where $R_\textrm{{eff}}$ is the galaxy's effective radius. Using annuli widths in units of $R_\textrm{{eff}}$ makes it possible to compare the radial profiles of galaxies with different sizes. The ratio of the major and minor axis and the position angle of the elliptical aperture are set using '$b/a$' ratio and position angle values from the value-added catalogue of \citet{Sanchez2022SDSS-IVGalaxies}. Fig.~\ref{oiiimaps} shows a source from the sample with the elliptical annuli used. We then estimate each annuli's pixel-weighted average $W_{80}$, using the routine described in \citet{Alban2024MappingCycleb}. Further, we only used galaxies with (a) at least 10 spaxels with a peak $S/N > 5$ detection., and (b) at least two annuli with 10\% area covered by spaxels with a peak $S/N > 5$. Although this drastically reduces the number of radio-AGN used later in our study, we use these criteria to obtain reliable radial profiles. The $S/N$ threshold of 5 is chosen to maximise the number of sources while avoiding contamination from low $S/N$ spectra. Changing the $S/N$ threshold to 3 or 10 changes the number of sources with \OIII radial profiles, but it does not alter our conclusions. Out of the 806 radio-AGN selected using LoTSS and FIRST, there are 378 sources with \OIII $W_{80}$ radial profiles. We remove 4 sources from this sample that show signs of mergers in their optical images and where the \OIII kinematics was affected by these interactions. Finally, we have a sample of 374 radio-AGN. From the control sample discussed above, we have 28 sources with \OIII radial profiles that we use for comparison with the AGN groups in Section~\ref{radialprof_ragn}.\par

The line width of \OIII profiles can provide insights into the impact of AGN feedback. However, the observed width of the profiles is determined by both - gravitational kinematics due to the motion of gas in the galaxy, and non-gravitational kinematics due to AGN and star formation-driven feedback. It is therefore crucial to assess the contribution of both to understand the extent to which \OIII gas is disturbed. We control for this first by constraining the sample to a certain stellar mass and stellar velocity dispersion limit, as described in the next section. Furthermore, in spatially resolved maps, \OIII profiles at the central region of the galaxy can also be broad due to the blending of narrow line profiles attributed to rotation (also depending on the inclination angle of the galaxy). This makes it hard to judge the impact of AGN solely with \OIII profile widths. We attempt to overcome this by normalising the \OIII $W_{80}$ with stellar $W_{80}$ values. This should (mostly) correct for the effects mentioned above since there is a broad correlation between \OIII and stellar velocity dispersions ($\sigma_{\star}$) observed in AGN host galaxies (e.g. \citealt{Nelson1996StellarBulge,Boroson2002DoesAGN,Sexton2020BayesianProperties}), and has been used before to trace AGN feedback (for e.g. \citealt{Woo2016,Ayubinia2023InvestigatingAGNs}). The stellar velocity dispersions would also be affected by the gravitational potential of the galaxy and the blending effect mentioned above. Therefore, any relative differences in the \OIII/ stellar $W_{80}$ ratios could be due to non-gravitational kinematics, tracing AGN or star formation-driven feedback. We construct these ratio maps as described below.\par

\begin{figure}
\centering
      \includegraphics[width=0.9\columnwidth]{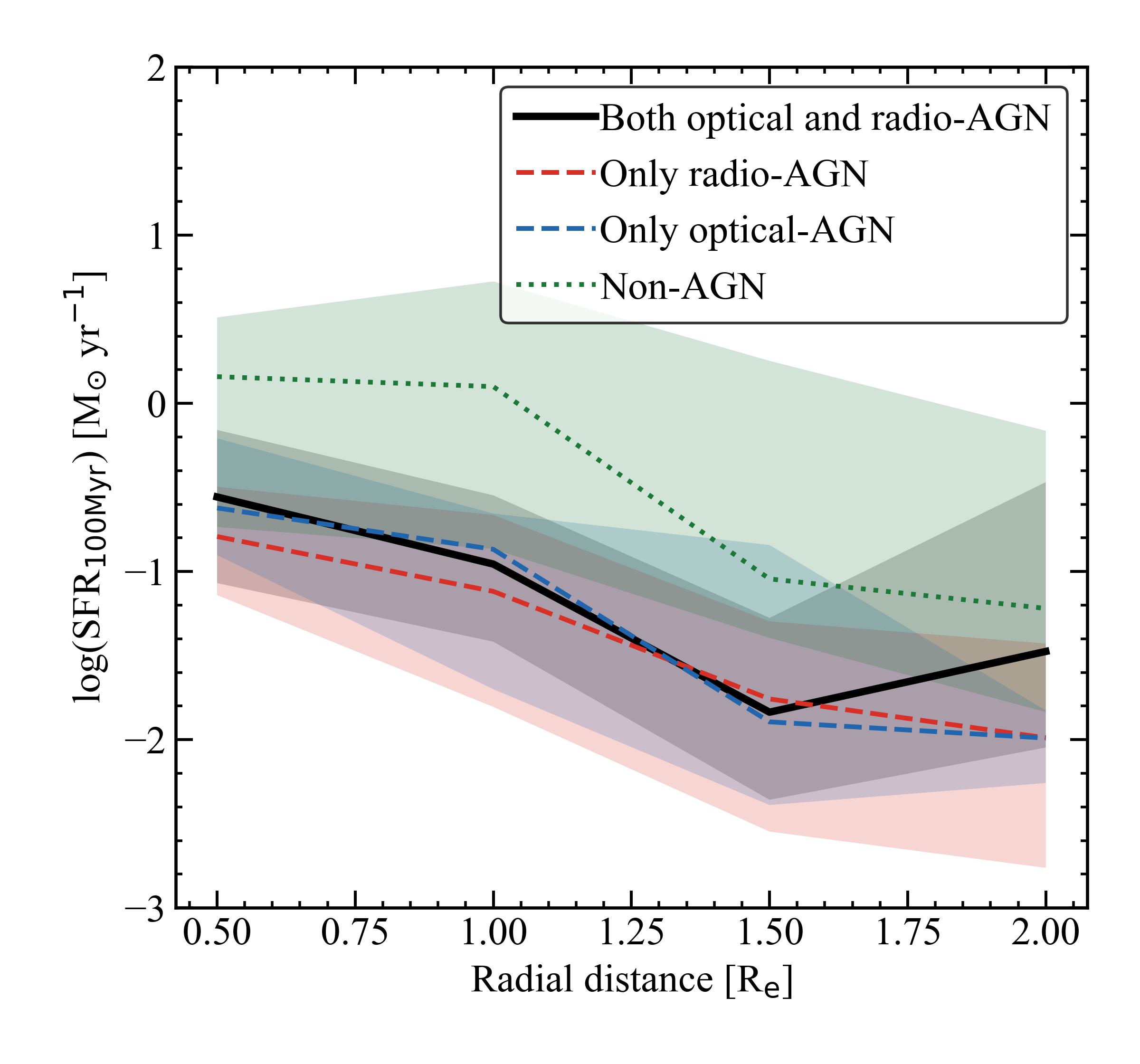}
      \caption{Radial profiles of recent (100 Myr) SFRs of the different source groups, discussed in Section~\ref{result ragn} and \ref{result luminosity}. The lines show the median SFRs and the shaded regions mark the 25th and 75th percentile values of the SFR distributions. SFRs for the non-AGN group are larger at every radial point than the AGN groups.}
    \label{sfrcomparison}
 \end{figure}

We first construct maps of $\sigma_{\star}$, using the \texttt{STELLAR\_SIGMA} extension in the \texttt{MAPS} files from the MaNGA DAP data products. We correct the stellar velocity dispersions using the \texttt{STELLAR\_SIGMACORR} extension included in the \texttt{MAPS} files, and as described in the online user manual. We then convert $\sigma_{\star}$ to a $W_{80}$ value, using a single Gaussian component conversion. Finally, we make the \OIII to stellar $W_{80}$ ratio maps (only for pixels with $S/N > 5$ \OIII detection) defined as $Q_{W80}\equiv \frac{\textrm{\OIII}\,W_{80}}{\mathrm{Stellar}\,W_{80}}$, and extract its radial profiles using the same approach outlined above. Examples of \q~maps are shown in Fig.~\ref{oiiimaps} for a radio-AGN source and Fig.~\ref{oiiimaps_nonagn} for a non-AGN source. An enhancement in the \q values in the central region can be seen in the radio-AGN source. These radial profiles are discussed further in Section~\ref{results}.

\section{Results}
\label{results}

In this section, we present the results for the \OIII kinematics of our radio-AGN sample and compare them to optical-AGN and the control sample of non-AGN sources selected above. We only use the sources with an \OIII detection for our analysis. For the most part, we will use the radial profiles of \width~and \q, to gauge the changes in these properties with distance from the galaxy centre, and disentangle the role of radio and \OIII luminosities. At the end of this section, we will also test for any relation between these properties and the radio-AGN alignment. The radio-AGN host galaxies are typically more massive ($M_{\star}>10^{11}M_{\odot}$) than optical-AGN and non-AGN galaxies. They also have larger average stellar velocity dispersions. We have therefore attempted to control for average $\sigma_{\star}$ (within an aperture of size \reff), $M_{\star}$ and \texttt{T-Type} of the different source groups as much as possible while keeping a sufficient number of sources in each group. Since the optical-AGN galaxies also have larger \OIII luminosities than radio-AGN galaxies and it can affect the \OIII radial profiles significantly, we also control for \OIII luminosities of the groups. Controlling for these properties and using normalised \oiiiwidth~(i.e. \q) on a pixel-by-pixel basis, accounts for any significant differences in the \OIII kinematics due to gravitational motion. However, this also reduces the number of sources in each group. The final number of sources in each group used for our analysis in this section, are summarised in Table~\ref{sourcegroups_table}. 

\subsection{Radial \OIII profiles of AGN groups }
\label{result ragn}

First, we compare the \oiiiwidth~and \q~radial profiles of the different AGN groups, selected in section~\ref{sample construction}, and shown in  Fig.~\ref{radialprof_ragn}. These radial profiles are assessed using violin plots, that help understand the distribution of the values at every radial point. The bottom panels of the figure show the p-values for a 2-sample KS test between these groups. Broadly, both \oiiiwidth and \q values show a decreasing trend with radial distance up to $r=1$\reff, with a flatter trend beyond this distance. Interestingly, we find that sources with radio+optical AGN have larger \oiiiwidth~than the group with either only radio or optical-AGN and non-AGN, up to $r=1.25$~\reff. But the p values show this difference is significant at $>$99\% significance only up to $r=0.5$~\reff when comparing to non-AGN group. However, normalising this for \stwidth, we find that the \q values are the largest for radio+optical AGN sources only up to $r=0.25$~\reff, at $>$99\% significance. Using the \q~values decreases the radial distance up to which we see a significant difference. However, we prefer to use the \q~values to judge the presence of disturbed gas since it accounts for gravitational motion and rotational line blending to some extent. This suggests that the differences in \oiiiwidth~seen at $r=1.25$\reff~were due to differences in the stellar velocity dispersion profiles, and not necessarily a sign of disturbed gas. At larger radial distances, the different group profiles seem to agree with each other, judging by the p-values.\par

It is worth noting that the \q~values at $r=0.25$ \reff~are largest for sources with radio+optical AGN, followed by optical and radio-AGN (which have roughly similar medians), and then non-AGN. These results show that when radio+optical AGN are present, the ionised gas in the central region could be most disturbed. Although there are significant number of \q~values below 1, the relatively larger values show that the \OIII line is broader in these sources than would be expected from gravitational motion only. Since the sample covers about three orders of magnitude in both $L_\mathrm{1.4GHz}$ and $L_{\textrm{\OIII}}$, it is possible that any luminosity-dependent differences are being washed out. We, therefore, perform the same analysis for sources above and below $L_\mathrm{1.4GHz}=10^{23}$\whz in the next section, to isolate any radio luminosity dependence.\par

The distributions of radio+optical AGN sources have a tail of large \oiiiwidth ($\gtrsim1000$\kms) and \q ($\gtrsim2$) values out to $r=1.25$\reff. Such high line widths likely denote disturbed ionised gas. However, since the radio-AGN in our sample are low-to-moderate radio luminosity systems, it is not entirely clear why these tails are mostly seen in radio+optical AGN. In moderate radio luminosity systems, jet-ISM interaction could drive fast outflows, which would explain the tail of this distribution. However, if the tail is mostly from low radio luminosity systems, it could also be an effect of the strong outflows shocking the surrounding gas and causing the radio emission. We discuss this further in section~\ref{result luminosity} and ~\ref{discussion}. Despite the tail of the distributions, we only find significant differences in the central small-scale region. \par

Beam smearing can affect the measured sizes of the kinematically disturbed \OIII region, and has been shown to overestimate the size in MaNGA data by \citet{Deconto-Machado2022IonisedCristino}. This could in part also explain the tail of the distributions that we observe out to large distances and discuss above. However, we only detect a significant difference on small scales ($r=0.25$\reff) compared to the MaNGA PSF (FWHM $\sim$2.5$\arcsec$). We therefore do not expect beam smearing to affect our results significantly.

As mentioned before, non-gravitational gas kinematics can also be due to feedback from star formation, which can disturb the ionised gas and drive outflows \citep{Heckman2015THEWINDS}. It is, therefore, important to ensure that any localised star formation is not driving any difference we observe in the \OIII kinematics is important. To compare the star formation rates (SFRs) of the groups, we use the radial profiles of SFRs measured by \citet{Riffel2023MappingSample}. These profiles were extracted using a similar approach but with annuli widths of 0.5\,\reff. Although their step size is twice ours, these profiles can still provide useful comparisons of the radial distribution of SFRs. In Fig.~\ref{sfrcomparison}, we compare the radial profiles of star-formation rates (SFRs) for our radio-AGN and non-AGN control samples. We find that the SFR radial profiles are in agreement for all AGN groups, whereas the non-AGN group has the largest values. Therefore, star formation does not drive the differences in the \OIII kinematics found between the AGN groups. Here we present only the SFRs averaged over the last 100 Myr, but we find the same results using SFRs averaged over the last 200 Myr.

\begin{table}[!ht]
\centering
\caption{Source groups.}
         \label{sourcegroups_table}
\renewcommand{\arraystretch}{1.15}
\setlength{\tabcolsep}{4pt}
\begin{tabular}{ccccc}
    \hlineB{3}
    \noalign{\vspace{0.05cm}}
    \hline
    \noalign{\smallskip}
     Group & All & $L_\mathrm{1.4GHz}$ & $L_\mathrm{1.4GHz}$ \\
      & & $10^{21}-10^{23}$ & $10^{23}-10^{25}$ \\
     
    \noalign{\smallskip}
    \hlineB{3}

    \noalign{\smallskip}
    \noalign{\smallskip}
  Radio+optical AGN & 32 & 20 & 12  \\
    \noalign{\smallskip}
  Radio-AGN & 184 & 142 & 42  \\
    \noalign{\smallskip}
  Optical-AGN & 22 & - & - \\ 
       \noalign{\smallskip}
  Non-AGN & 28 & - & - \\
    \noalign{\smallskip}
    \noalign{\smallskip}
   
    \hline
    \noalign{\vspace{0.05cm}}  
    \hlineB{3}
    \end{tabular}
    \flushleft
    Note. Number of sources in different source groups discussed in Section~\ref{results}, after matching the groups in host galaxy properties. These only include sources with an \OIII detection in the MaNGA data.     
\end{table}

 \begin{figure*}[!ht]
    \centering
    \begin{subfigure}{1.7\columnwidth}
      \includegraphics[width=\columnwidth]{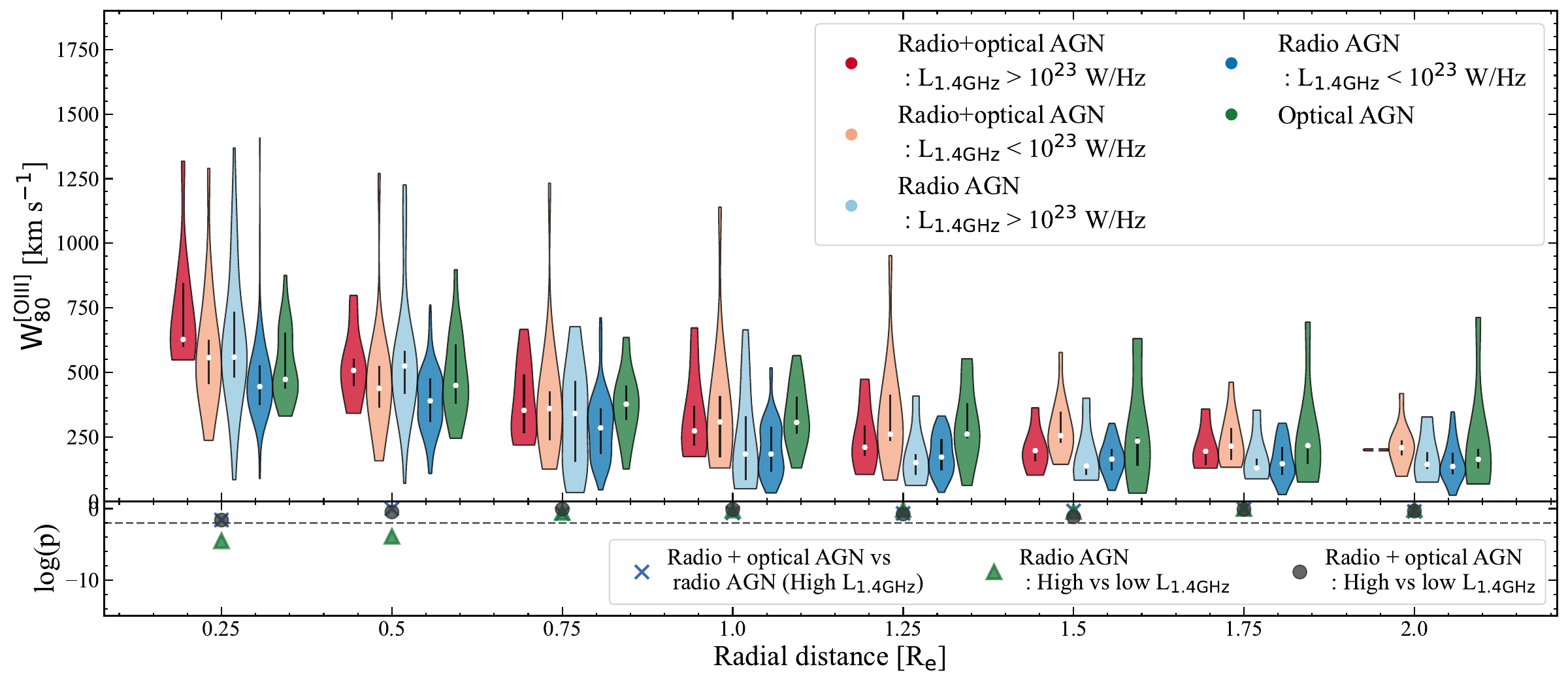}
      \caption{Radial profiles of \oiiiwidth for $L_\mathrm{1.4GHz}>10^{23}$\whz radio-AGN.}
      \label{radialprof_ragnrlum_oiiiw80}
    \end{subfigure}
    \begin{subfigure}{1.7\columnwidth}
      \includegraphics[width=\columnwidth]{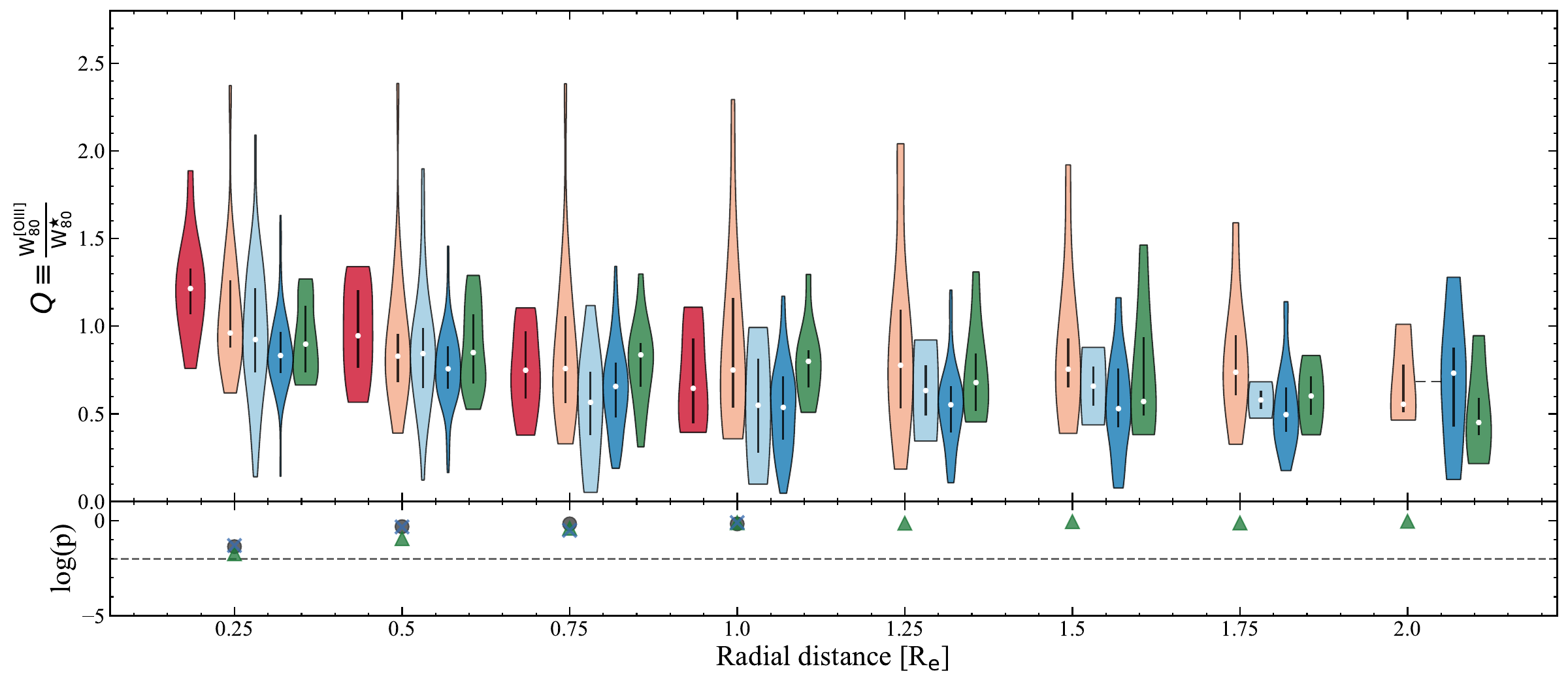}
      \caption{Radial profiles of \q~for $L_\mathrm{1.4GHz}>10^{23}$\whz radio-AGN.}
      \label{radialprof_ragnrlum_q}
    \end{subfigure}
    
    \begin{subfigure}{1\columnwidth}
      \includegraphics[width=\columnwidth]{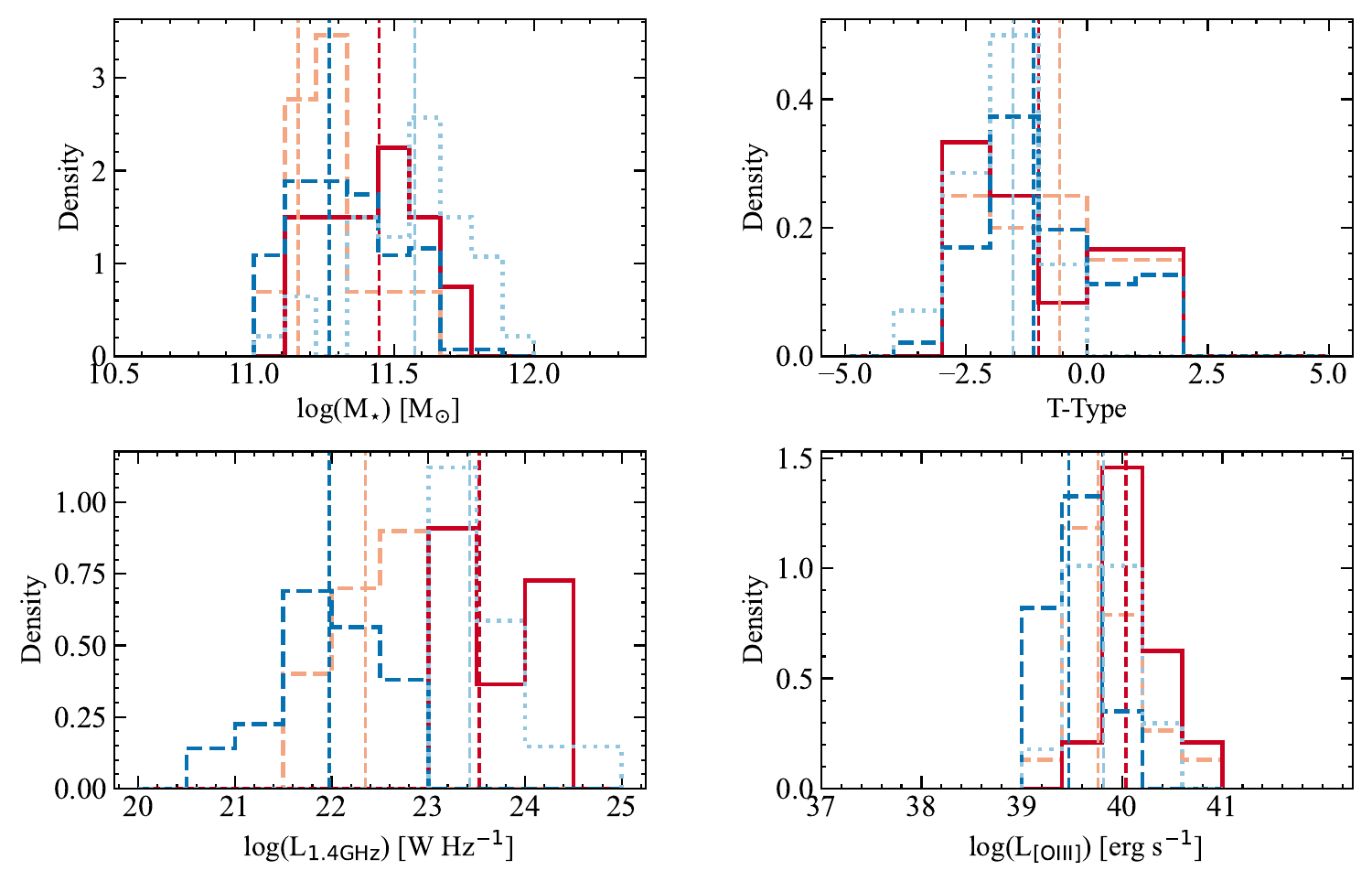}
      \caption{Host galaxy properties for the groups}
      \label{ragnrlum_prop}
    \end{subfigure}   
    \caption{Same radial profiles in (a) and (b) as in Fig.~\ref{radialprof_ragn} but for high and low $L_\mathrm{1.4GHz}$ radio+optical AGN and radio-AGN sources. The colours in panel (c) represent the same sources as in the top panels.}
    \label{radialprof_ragnlum}
 \end{figure*}

\subsection{Dependence on radio luminosity}
\label{result luminosity}

The total mechanical energy output of a radio-AGN is correlated to $L_\mathrm{1.4GHz}$ (e.g. \citealt{Cavagnolo2010,McNamara2012}). Therefore, investigating the dependence of \OIII kinematics on radio luminosity can help understand the relation between feedback, and the total mechanical energy emitted. Using the spatially resolved maps we can determine if the scales on which \OIII is disturbed change with $L_\mathrm{1.4GHz}$. Since the radio-AGN sample also has some optical-AGN, we handle the radio-AGN sources with and without an optical-AGN separately, as done above. We then split the sample into groups of radio luminosity at $L_\mathrm{1.4GHz}=10^{23}$\whz. Low $L_\mathrm{1.4GHz}$ sources cover a luminosity range of $10^{21}-10^{23}$\whz and high $L_\mathrm{1.4GHz}$ sources from $10^{23}-10^{25}$\whz, although most sources in the high $L_\mathrm{1.4GHz}$ groups are between $10^{23}-10^{24}$\whz. In the group of radio+optical AGN, this gives 20 low $L_\mathrm{1.4GHz}$ sources and 12 high $L_\mathrm{1.4GHz}$ sources. In the group of radio-AGN, this gives 142 low $L_\mathrm{1.4GHz}$ sources and 42 high $L_\mathrm{1.4GHz}$ sources (summarised in Table~\ref{sourcegroups_table}). \par

The radial profiles of these four groups are shown in Fig.~\ref{radialprof_ragnlum}. We find that the median values of \oiiiwidth and \q~are largest for radio+optical AGN with high $L_\mathrm{1.4GHz}$, within $r=0.25$\reff \footnote{The \q~radial profiles of this group have very few ($<5$) points beyond $r=1$\,\reff, therefore the distributions are not reliable and are not shown here}. However, this difference is only marginally significant when compared to radio+optical AGN with low $L_\mathrm{1.4GHz}$. Similarly, radio-AGN with high $L_\mathrm{1.4GHz}$ show larger \oiiiwidth and \q~up to $r=0.5$\reff, but it is only marginally significant till $r=0.25$\reff. Judging by the p values of the distributions beyond $r=0.5$\reff, we find that distributions for all groups agree with each other. This shows that the presence of a moderately powerful radio+optical AGN disturbs the \OIII gas most, although the spatial extent at which gas is disturbed does not increase at higher $L_\mathrm{1.4GHz}$.\par

Looking at the \q~radial profiles of the radio+optical AGN with low $L_\mathrm{1.4GHz}$ in Fig.~\ref{radialprof_ragnrlum_q}, it is interesting to note that they have a tail of high \q values out to large radial distances ($r=1.75$\reff). However, this tail is not visible in the radio+optical AGN with high $L_\mathrm{1.4GHz}$. We described this tail in the previous section. Given that its most prominent in the low $L_\mathrm{1.4GHz}$ radio+optical AGN, a likely explanation for this could be that the radio emission observed in these sources is due to shocks driven by the radiation from the optical-AGN, and is, therefore, an effect of the disturbed ionised gas, and not the cause. This shock interpretation is discussed more in section~\ref{discussion}.\par

\section{Discussion}
\label{discussion}

In this paper, we have studied the impact of low-to-moderate luminosity radio and optical-AGN on the \OIII gas in the local Universe up to $z\approx0.15$. We selected a sample of 806 nearby radio-AGN from the MaNGA catalogue, using a combination of LoTSS and FIRST surveys, out of which 378 have an \OIII detection. We then controlled for host galaxy properties and finally used a sample of 32 radio+optical AGN, 184 radio-AGN, 22 optical AGN and 28 non AGN sources for our analysis. Constructing such groups allowed us to systematically study the feedback from radio and optical AGN selected from the same survey. Using spatially resolved maps of \OIII spectra, we constructed radial profiles of \oiiiwidth~and \q, to study the impact on the warm ionised gas. Our sample covers a radio luminosity range of $L_\mathrm{1.4GHz}\approx10^{21}-10^{24}$\,\whz, including systems that are traditionally called radio-loud and radio-quiet AGN. \par

\subsection{Impact of radio-AGN on \OIII}
Comparing the \oiiiwidth~and \q~radial profiles of different AGN groups from our sample, allows us to determine the presence of disturbed \OIII and its relation to the radio and optical-AGN. Comparing the \q~profiles of all sources in Fig.~\ref{radialprof_ragn}, we find a statistically significant difference at $r=0.25$\,\reff, between sources that have radio+optical AGN, and sources that have only either radio or optical-AGN. We also find that when radio+optical AGN are present, more than half the sources have \q$>1$ at $r=0.25$\,\reff. This shows that \OIII is more disturbed than when radio+optical AGN are present, compared to when only either one is present. The radial distance of $r=0.25$\reff, corresponds to a physical radial distance range of $\sim0.5-5.9$\,kpc for our sample, and 1.9\,kpc at the median redshift. We propose that the warm ionised gas in nearby AGN host galaxies is disturbed on compact scales. Similar results for the compact nature of disturbed ionised gas have been found before (for example  \citealt{Tadhunter2018QuantifyingObservations,Holden2025NoSmearing}).\par

Beyond this point, we find that the \q~radial profiles of all AGN groups are largely similar, except the high values that create the tails seen in the distributions of radio+optical AGN. Although similar trends can be seen in the \oiiiwidth~radial profiles of the different groups, the differences between them are suppressed when normalising them with \stwidth. This shows that the differences in the \OIII kinematics can also be attributed to locally different gravitational motions, and a difference in \oiiiwidth~does not necessarily imply more or less disturbed \OIII kinematics. This highlights the advantage and necessity of using \q values for comparing different sources. \par

In the radio+optical AGN group, both jets and shocks can be responsible for the observed radio emission, as discussed in section~\ref{selecting radio-AGN}.  In the case of jets, the presence of more disturbed \OIII in radio+optical AGN would point to a co-active role of jets and radiation, such that when both are present, the impact on warm ionised gas kinematics is the strongest. In the case of radiative wind-driven shocks, this would mean that the disturbed \OIII is essentially tracing sources with strong shocks, that cause the radio emission. Therefore, the radio-emission would not be the cause but the effect of the feedback on the host galaxy.\par

Differentiating between the two cases would ideally require high-resolution radio imaging and radio spectral analysis. Lacking this information for the sample, splitting the radio+optical AGN sample into high and low $L_\mathrm{1.4GHz}$ sources sheds some light on this. We find that the trend we discussed above, with more than half of these sources having \q$>1$ at $r=0.25$\,\reff, is driven by the high luminosity sample, as can be seen in Fig.~\ref{radialprof_ragnlum}. These systems are more likely to have their radio emission dominated by jets, as can be seen by their positions with respect to the models in Fig.~\ref{source prop}. Therefore, we propose that in these systems, the radio emission is dominated by jets and the signatures of disturbed ionised gas observed on small scale are likely due to jet-ISM interaction.
\par
However, the prominent tail of large \oiiiwidth and \q~values in the low $L_\mathrm{1.4GHz}$ sources likely point towards a shock origin of radio emission in these systems (although beam smearing also contributes to this, as mentioned before). Indeed, these systems also fall on or below the \citep{Nims2015ObservationalNuclei} models shown in Fig.~\ref{source prop}. Therefore, it is possible that at these low radio luminosities, the high \OIII~line widths are a selection effect, as selecting radio-AGN means essentially selecting for sources with strong shocks. Overall, this highlights the role of both sources of radio emission in understanding feedback in low to moderate radio luminosity systems. \par

Our results reinforce the positive correlation between disturbed \OIII kinematics and the presence of radio emission in AGN host galaxies, which has also been observed before with SDSS single fibre spectra that cover the central 3\arcsec of the galaxies (e.g. \citealt{Mullaney2013,Woo2016,Molyneux2019,Kukreti2023,Kukreti2024}). Recently, using a sample of $\sim$5\,700 radio-AGN and SDSS spectra, \citet{Kukreti2024} found a positive correlation between the \OIII line-widths and $L_\mathrm{1.4GHz}$, with more disturbed gas sources with $L_\mathrm{1.4GHz}>10^{23}$\,\whz. Using a large sample of $\approx24\,000$ type 1 and 2 AGN, \citet{Mullaney2013} also found that the width of \OIII profiles was larger for AGN with a radio detection, peaking between $10^{23}-10^{25}$\,\whz. However, the sources in this study were not selected to be radio-AGN, and the shock origin of radio-emission could be quite significant contributor in this sample. Although the single fibre spectra used in these studies only cover the innermost region of our galaxies as mentioned above, our results are still in agreement with these studies over that region. The 0.25\,\reff~point up to which we detect disturbed \OIII covers the same size of central region as the SDSS single fibre studies at these redshifts. However, the spatially resolved MaNGA IFU data allows us to investigate the impact on larger scales, where we find no significant differences in the impact of various low-to-moderate luminosity AGN groups. \par

\subsection{The simultaneous impact of jets and radiation}

Separating sources with radio+optical AGN from those with either only radio or optical AGN allows us to perform a comparative analysis of the role of jets and radiation in low-to-moderate luminosity AGN. This is illustrated in Fig.~\ref{radialprof_ragnlum}, which shows the \oiiiwidth~and \q~radial profiles for different $L_\mathrm{1.4GHz}$ of radio-AGN sources, with and without an optical-AGN. When radio+optical AGN are present in a source, we find that \OIII is more likely to be disturbed on small scales, when the radio-AGN has $L_\mathrm{1.4GHz}>10^{23}$\,\whz than when it is less luminous.
\par
Even though optical-AGN of similar \OIII luminosities are present in both $L_\mathrm{1.4GHz}$ groups, the presence of more powerful jets correlates with relatively more disturbed \OIII, judging by the difference in \q~values. These jets are likely the source of radio emission in the high $L_\mathrm{1.4GHz}$ sources, as mentioned above. This group of radio+optical AGN has more disturbed \OIII sources at 0.25\reff~than their low $L_\mathrm{1.4GHz}$ counterparts and the optical-AGN group. This shows that when moderately powerful radio-AGN and optical-AGN are present in a source simultaneously, the gas is most likely to be disturbed. Therefore, jets and radiation in these systems seem to be acting in a manner where the both enhance the impact of each other on the surrounding ionised gas kinematics. When radiation pressure from the AGN is present along with the jets, the AGN is more effective in disturbing the \OIII kinematics than it would be if it only had jets (and vice versa). Comparing only points with \q$>1$, we see that the impact is strongest for the group with radio+optical AGN and radio-AGN with high $L_\mathrm{1.4GHz}$. We remind the reader that sources with $L_\mathrm{1.4GHz}>10^{23}$\,\whz are still moderate luminosity sources, with typical luminosities between $10^{23}-10^{24}$\,\whz.  \par

Our results show that the strongest impact of jets in radio+optical AGN systems is limited to the central region of the galaxies. Comparing the radio sizes and optical host galaxy sizes sheds some more light on this trend. The ratio between the radio sizes of our sources, which we take from the LoTSS value-added catalogue of \citet{Hardcastle2023TheRelease}, and the effective radii of the host galaxies, has a median value of 0.5. This means that the radio emission from the AGN is typically on much smaller scales than the host galaxy size. This explains why the impact we see on \OIII in the $L_\mathrm{1.4GHz}>10^{23}$\,\whz radio-AGN groups, is limited to the central region. \par

Combining this with the results discussed above, we propose a picture where jet and radiation-driven feedback are active simultaneously in moderate luminosity radio-AGN host galaxies. The ionised gas appears to be impacted by both, as can be seen in the highly disturbed radial profiles of sources that have both AGN. This shows that feedback on ionised gas in AGN selected to have radio jets, is not necessarily only driven by the jets. The presence of radiation from the AGN makes it more likely for the gas to be kinematically disturbed in radio-AGN host galaxies, compared to when only jets are present. Further, comparing the radio-AGN groups with and without an optical-AGN suggests that gas clouds perhaps pushed to high velocities by the jets are driven to even higher velocities by the impact of radiation, and vice versa.\par

\subsection{Feedback in low-to-moderate luminosity radio-AGN}

Feedback on \OIII in radio-AGN host galaxies has also been studied by \citet{Alban2024MappingCycleb}, and we have used the same routine to extract the radial \width~profiles for our study. In their analysis of radio-AGN, \citet{Alban2024MappingCycleb} find that radio-AGN show large \width~values at large \reff~in comparison to broad-line and optical AGN, matched in other host galaxy properties. We find that \OIII is most disturbed, in terms of gas velocity and proportion of sources disturbed, in the central region when radio+optical AGN are present. The impact is relatively milder when either only radio or optical-AGN are present. This is qualitatively in agreement with the results of \citet{Alban2024MappingCycleb}, however, the differences we find in the impact on \OIII are limited to a radial distance of 0.25\reff. The differences in the \OIII line widths reported by \citet{Alban2024MappingCycleb} beyond this distance, could be due to the low luminosity radio+optical AGN that we have in our sample. \par

It is also worth noting that these differences could be due to the differences in the selection techniques, and the host galaxies of the radio-AGN selected. The radio-AGN we select are predominantly hosted by passive early-type galaxies, whereas theirs are equally distributed among early and late-type galaxies. Since early-type galaxies are dispersion-dominated and passive in terms of star formation, their \q~and SFR are lower than late-type galaxies. This explains the differences observed in the two samples at larger radii. Indeed controlling for early-type (or late-type) galaxies brings the radial profiles in agreement with each other.\par

Our results propose a picture where the impact of low-to-moderate luminosity radio-AGN is strongest on the ionised gas in the central $0.25$\,\reff~region of galaxies in the local Universe. The lifetime of a radio-AGN phase of $\sim$\,$10-100$\,Myr is significantly smaller than the galaxy's lifetime of a few Gyr (see \citealt{Morganti2017b} for a review). Therefore, a single phase of activity is unlikely to impact star-formation significantly over the galaxy's lifetime. Indeed, many radio-AGN with multiple epochs of activity have been detected (e.g. \citealt{Sridhar2020,Brienza2020,Kukreti2022a}), pushing the need to understand the cumulative impact of multiple AGN phases on the galaxy. If every phase of activity had the strongest impact on the gas in the central regions, we would expect to see less star formation in these regions compared to non-AGN host galaxies. Evidence for such inside-out quenching has been found in MaNGA galaxies, albeit in AGN selected using emission line ratios (e.g. \citealt{Bluck2020HowSurvey,Lammers2023ActiveRates,Bertemes2022}). AGN feedback over multiple phases of activity could have suppressed the star formation in the central regions of these galaxies. Although we study the warm ionised phase which is not the fuel for star formation, our results provide evidence supporting this scenario. Further studies of feedback in restarted radio-AGN on the molecular gas phases (fuel for star formation) would be required to test these scenarios. We plan to conduct such studies in the future. \par

\section{Summary}

We have selected a sample of radio-AGN using LoTSS and FIRST surveys and combined them with MaNGA to obtain spatially resolved spectra for a subsample over a redshift range of $z=0.01-0.15$. The radio-AGN are selected to have larger radio emission than what is expected from star-formation, and are low-to-moderate radio and \OIII luminosity sources. We assess the impact of radio and optical-AGN on the \OIII kinematics in these sources, to disentangle the role of jets and radiation in driving feedback. Our main finding is that when radio+optical AGN are present, sources are significantly more likely to have disturbed \OIII up to a radial distance of 0.25\,\reff, than when either only radio or optical-AGN are present. This shows that when both jets and radiation are present in a system, the AGN have the strongest impact on the surrounding warm ionised gas. This relation is dependent on radio luminosity, and high radio luminosity sources are more likely to have more disturbed gas in the central region. We note that in the low radio luminosity radio+optical AGN, the observed radio emission could be due to wind-driven shocks instead of jets. However, this doesn't affect our results, which are mainly driven by the higher radio luminosity AGN. Finally, we find that any differences in the impact on \OIII are only visible up to a radial distance of 0.25\,\reff. We find no evidence of a widely occurring large-scale impact of moderate luminosity jets on the warm ionised gas in these galaxies.

\begin{acknowledgements} 
We thank the anonymous referee for the feedback and suggestions, that improved the paper. D.W. acknowledges support through an Emmy Noether Grant of the German Research Foundation, a stipend by the Daimler and Benz
Foundation and a Verbundforschung grant by the German Space Agency. BDO acknowledges the support from the Coordena{\c c}{\~a}o de Aperfei{\c c}oamento de Pessoal de N\'ivel Superior (CAPES-Brasil, 88887.985730/2024-00).
 This project makes use of the MaNGA-Pipe3D dataproducts. We thank the IA-UNAM MaNGA team for creating this catalogue,
and the Conacyt Project CB-285080 for supporting them. Funding for the Sloan
Digital Sky Survey IV has been provided by the Alfred P. Sloan Foundation,
the U.S. Department of Energy Office of Science, and the Participating Institutions. SDSS-IV acknowledges support and resources from the Center for High-Performance Computing at the University of Utah. The SDSS website is www.sdss.org. 
\end{acknowledgements}
  \bibliographystyle{aa-copy} 
  \bibliography{References}

\begin{appendix}

\section{Selecting radio-AGN from FIRST}
\label{selecting radio agn first}

\begin{table}[!ht]
\centering
\caption{Diagnostic combinations for FIRST sources}
         \label{diagnosticcomb_first}
\renewcommand{\arraystretch}{1.15}
\setlength{\tabcolsep}{4pt}
\begin{tabular}{cccccc}
    \hlineB{3}
    \noalign{\vspace{0.05cm}}
    \hline
    \noalign{\smallskip}
     $D_\mathrm{{n}}(4000)$  & BPT  & $L_{\textrm{H}\alpha}$  & WISE & Number &  Final \\
     vs $L_\mathrm{1.4GHz}$/$M_{\star}$ &   & vs $L_\mathrm{1.4GHz}$ & col-col &  & class\\
     
    \noalign{\smallskip}
    \hlineB{3}

    \noalign{\smallskip}
    \noalign{\smallskip}
  
AGN & AGN & AGN & AGN & 46 & AGN\\
AGN & AGN & AGN & Uncl. & 17 & AGN\\
AGN & Int. & AGN & AGN & 21 & AGN\\
AGN & Uncl. & AGN & AGN & 27 & AGN\\
AGN & Uncl. & AGN & Uncl. & 15 & AGN\\
AGN & Uncl. & Uncl. & Uncl. & 21 & AGN\\
Int. & AGN & AGN & AGN & 15 & AGN\\
Int. & AGN & Int. & SF & 16 & SF\\
Int. & AGN & Int. & Uncl. & 14 & AGN\\
Int. & AGN & SF & AGN & 24 & SF\\
Int. & Int. & AGN & AGN & 10 & AGN\\
Int. & Int. & Int. & AGN & 19 & AGN\\
Int. & Int. & Int. & SF & 14 & SF\\
Int. & Int. & SF & AGN & 10 & AGN\\
Int. & Uncl. & AGN & AGN & 12 & AGN\\
Int. & Uncl. & Uncl. & AGN & 32 & AGN\\
Int. & Uncl. & Uncl. & Uncl. & 13 & AGN\\
AGN & Uncl. & Uncl. & AGN & 43 & AGN\\ 
Int. & AGN & Int. & AGN & 61 & AGN\\
SF & AGN & Int. & SF & 22 & SF\\ 
SF & AGN & SF & SF & 40 & SF\\ 
SF & Int. & Int. & SF & 49 & SF\\ 
SF & Int. & SF & SF & 99 & SF\\ 
SF & SF & Int. & SF & 65 & SF\\ 
SF & SF & SF & SF & 254 & SF\\ 
SF & SF & SF & Uncl. & 15 & SF\\ 
Uncl. & Uncl. & Uncl. & Uncl. & 19 & Uncl.\\ 

    \noalign{\smallskip}
    \noalign{\smallskip}
   
    \hline
    \noalign{\vspace{0.05cm}}  
    \hlineB{3}
    \end{tabular}
    \flushleft
    Note. Combinations for classification of a source using the diagnostic diagrams discussed in Section~\ref{selecting radio-AGN}, and the final classification assigned. Only groups with more than 40 sources are shown here.   
\end{table}

\begin{table}[!ht]
\centering
\caption{Classification of FIRST cross-matched sources}
         \label{diagnostictable_first}
\renewcommand{\arraystretch}{1.15}
\setlength{\tabcolsep}{4pt}
\begin{tabular}{ccccc}
    \hlineB{3}
    \noalign{\vspace{0.05cm}}
    \hline
    \noalign{\smallskip}
     Diagnostic & AGN & Int. & SF & Uncl.\\
     & \multicolumn{4}{c} {(No. of overall radio-AGN)} \\ 
     
    \noalign{\smallskip}
    \hlineB{3}

    \noalign{\smallskip}
    \noalign{\smallskip}
  
  $D_\mathrm{{n}}(4000)$ vs $L_\mathrm{1.4GHz}$/$M_{\star}$ & 261 & 277 & 611 & 6  \\ 
   & (260) & (227) & (25) & -  \\ 
    \noalign{\smallskip}
  $L_{\textrm{H}\alpha}$ vs $L_\mathrm{1.4GHz}$ & 288 & 299 & 457 & 111 \\
   & (270) & (124) & (11) & (107)  \\ 
    \noalign{\smallskip}
  BPT & 360 & 278 & 324 & 200 \\
   & (233) & (80) & - & (191) \\
       \noalign{\smallskip}
  WISE col-col & 440 & - & 554 & 161 \\
   & (374) & - & (17) & (121)  \\ 
    \noalign{\smallskip}
    \noalign{\smallskip}
   
    \hline
    \noalign{\vspace{0.05cm}}  
    \hlineB{3}
    \end{tabular}
    \flushleft
    Note. Number of sources classified by each diagnostic discussed in Section~\ref{selecting radio-AGN}. The different classes are - AGN, intermediate (Int.), star-forming (SF) and unclassified (Uncl.). The numbers in brackets are sources from each group classified as radio-AGN after combining the four diagnostics..    
\end{table}

\begin{figure*}
  \centering
      \includegraphics[width=1.6\columnwidth]{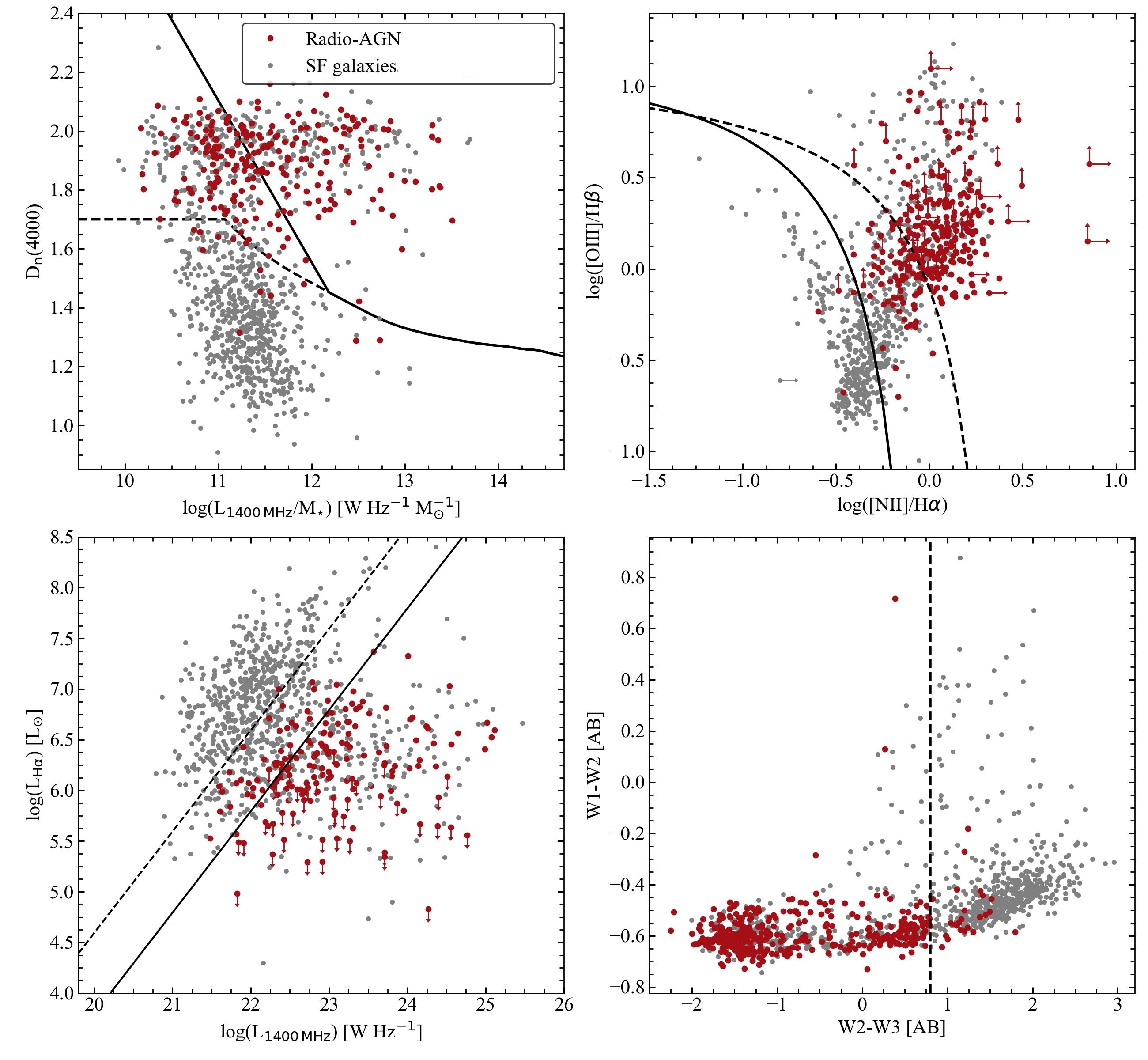}
      \caption{Same plot as in Fig.~\ref{diagnostic_lotss}, but for selecting radio-AGN using FIRST data.}
    \label{diagnostic_first}
\end{figure*}

\begin{figure*}
    \centering
    \includegraphics[width=1.8\columnwidth]{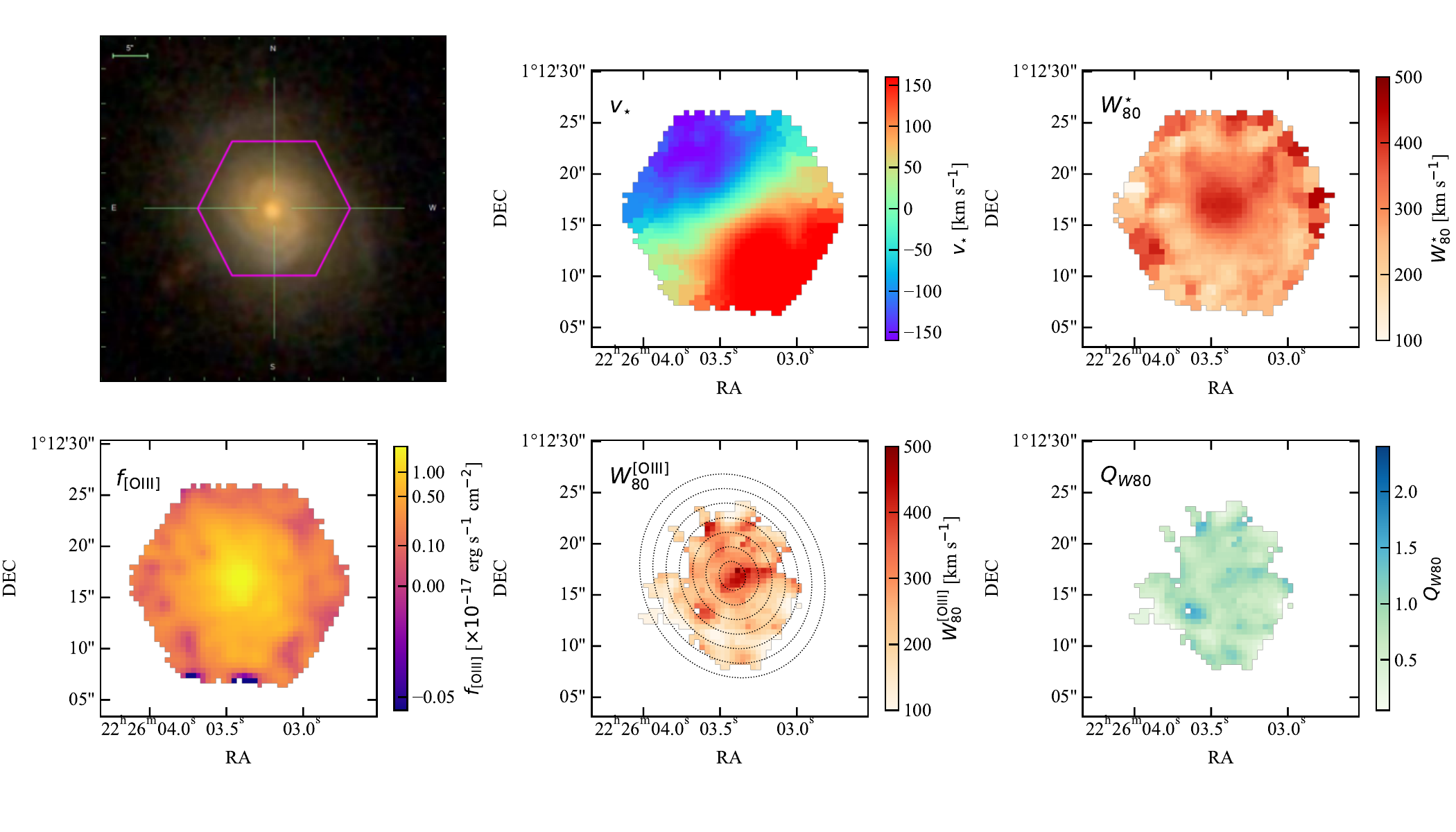}
    \caption{Same as Fig.~\ref{oiiimaps} but for a non-AGN galaxy (plateifu=12068-6104).}
    \label{oiiimaps_nonagn}
 \end{figure*}

\end{appendix}

\end{document}